%% file: manuscript.tex
\pdfoutput=1
\documentclass[manuscript,screen,nonacm]{acmart}
\usepackage{subcaption}
\usepackage{array}
\usepackage{listings}
\usepackage{graphicx}

\AtBeginDocument{%
  }

\begin{document}

\title{Adaptive Authorization as a Second Line of Defense:\texorpdfstring{\\}{ } Mitigating Residual Risks in Adaptive Authentication}

\author{Alkabashi Alnour}
\email{alkabashi.alnour@ucdconnect.ie}
\affiliation{%
  \institution{University College Dublin and Lero}
  \city{Dublin}
  \country{Ireland}
}

\author{Bashar Nuseibeh}
\affiliation{%
  \institution{City St George’s, University of London}
  \city{London}
  \country{UK}}
\email{Bashar.Nuseibeh@citystgeorges.ac.uk}

\author{Liliana Pasquale}
\affiliation{%
  \institution{University College Dublin and Lero}
  \city{Dublin}
  \country{Ireland}
}
\email{liliana.pasquale@ucd.ie}

\author{Alzubair Hassan}
\email{alzubair.hassan@ucd.ie}
\affiliation{%
  \institution{University College Dublin and Lero}
  \city{Dublin}
  \country{Ireland}
}

\renewcommand{\shortauthors}{Alnour et al.}
\renewcommand{\shorttitle}{Adaptive Authorization as a Second Line of Defense}

\begin{abstract}
Adaptive authentication selects authentication methods according to changing contextual conditions to balance security, usability, and performance. However, risks of impersonation, replay, and session hijacking can persist after authentication, and existing approaches provide limited support for using these \emph{residual risk}s to guide authorization decisions. We propose an approach that extends adaptive authentication with adaptive authorization as a second line of defense. The approach quantifies the residual risk associated with the selected authentication method and identifies authorization restrictions that limit the potential impact of compromise while preserving critical operations. A formal decision model combines operation criticality, dependencies between operations and assets, and fine-grained access rights with a penalty model that quantifies functionality loss from tightening access rights or disabling operations. We encode the decision problem in SMT-LIB and use Z3 to generate feasible authorization configurations with different trade-offs between residual risk reduction and functionality preservation. We evaluate the approach across six scenarios in healthcare and industrial control systems, comparing it with representative authorization baselines and conducting additional analyses of parameter sensitivity and computational overhead. The approach identifies configurations that reduce both attack-specific and overall modeled residual risk while keeping all necessary operations enabled. It achieves more favorable risk–functionality trade-offs than least-privilege role-based access control in all six scenarios and attribute-based access control in five scenarios. By making these trade-offs explicit, the approach supports administrators in selecting authorization configurations that reflect operational priorities and organizational risk appetite.
\end{abstract}


\maketitle

\input{Introduction}
\input{Scenarios}
\input{Approach}
\input{Evaluation}
\input{Related_Work}
\input{Conclusion}

\begin{acks}
This work was partially supported by Research Ireland grant 13/RC/2094\_P2.
\end{acks}

\end{document}

%% file: Introduction.tex
\section{Introduction}

Adaptive authentication selects an authentication method at runtime according to contextual factors such as device, network, location, and user behavior, balancing security against usability and performance requirements~\cite{hassan_engineering_2021,arias-cabarcos_survey_2020}. As adaptive authentication is increasingly considered in domains such as healthcare \cite{imprivata_healthcare_2026} and the Internet of Vehicles \cite{10.1145/3828751}, understanding what risks remain after authentication has become more important.   Highly usable methods such as fingerprint recognition remain exposed to spoofing and replay attacks~\cite{arias-cabarcos_survey_2020}, and even multi-factor authentication (MFA) can be bypassed in practice. In the 2022 Uber breach, an attacker with a contractor's stolen password repeatedly triggered MFA approval requests until the contractor accepted one, then gained access to other employee accounts and, ultimately, elevated permissions to internal tools~\cite{uber_security_2022}. In 2023, attackers who breached Okta's customer support system extracted session tokens from customer-uploaded files and used them to hijack those customers' legitimate, already-authenticated sessions~\cite{okta_support_2023}. In both cases, the damage the attackers could cause depended less on the strength of authentication than on what the compromised account or session was authorized to do. This raises the question that motivates this paper: once an authentication method has been applied, what risk remains, and how should the system respond to it?

We refer to the risk that remains after the selected authentication method is applied as \textit{residual risk}. Existing adaptive authentication approaches (e.g., \cite{hassan_engineering_2021,arias-cabarcos_survey_2020,papaioannou_risk-based_2021}) primarily focus on selecting and applying authentication methods, with limited support for mitigating the risk that remains. Consider a physician who logs in from an ambulance during an emergency. An adaptive authentication system may prioritize usability and select a smartcard alone, while the session may remain exposed to replay and session hijacking over an unsecured Wi-Fi connection~\cite{fi15040146}, putting patient records and diagnoses at risk. Requiring stronger or additional authentication may be impractical or insufficient: additional steps can delay urgent care, the available device may not support other authentication factors, and successful authentication does not fully prevent subsequent session hijacking. The system can, however, adapt the permissions granted to the authenticated session. By restricting accessible assets and permitted actions, adaptive authorization can limit the potential consequences of compromise and provide a second line of defense.

Step-up authentication allows a resource to require stronger or more recent authentication before granting a sensitive request~\cite{rfc9470}, while digital identity guidelines define authentication assurance levels to support stronger requirements for higher-risk access~\cite{nist_sp800_63b4}. Zero-trust architectures authorize access to resources on a per-session basis~\cite{nist_sp800_207}. Risk-aware access control weighs trust, including authentication strength, against the risk associated with an access request~\cite{debruyne_balancing_2015}, and self-adaptive authorization frameworks reconfigure access-control policies at runtime~\cite{bailey_self-adaptive_2011}. These mechanisms, however, either ask for more authentication, which is often infeasible in the situations described above, or derive access decisions from authentication strength and context through predefined policies or per-request rules. To the best of our knowledge, none of them quantifies the residual risk left by the specific method that adaptive authentication selects in the current context, or systematically explores how much functionality must be restricted, and where, to reduce it.

This paper extends adaptive authentication with adaptive authorization as a second line of defense against residual risk. Its key idea is that every attack we consider, whether impersonation, replay, or session hijacking, ends in an authenticated session whose capabilities are bounded by the authorization configuration. Residual risk can therefore be reduced not only by making a successful attack less likely, but also by limiting what a compromised session could do. After adaptive authentication selects a method, our approach quantifies the residual risk that the method leaves and characterizes authorization configurations that reduce this risk while preserving access to the assets required by the most critical operations. To this end, it classifies operations by criticality into \textit{Must-Have}, \textit{Nice-to-Have}, and \textit{Optional}; maps each operation to the assets it uses and the access rights (Read, Write, Update, Delete) it requires on them; and introduces a penalty model that captures the functionality lost by tightening access rights or disabling operations. The resulting decision problem is encoded in SMT-LIB and solved with Z3 \cite{10.1007/978-3-540-78800-3_24}. On the authentication side, the approach only needs, for each modeled attack, the likelihood that remains after the selected method is applied, so it can be combined with any adaptive authentication mechanism that provides these likelihoods. In this paper, we instantiate it with our requirement-based adaptive authentication framework~\cite{10.1145/3828751}, which selects methods by trading off security, usability, and performance requirements in the current context.

Because residual risk and functionality loss have no common scale, the approach does not prescribe a single policy. Instead, it exposes the trade-off as a set of recommended configurations, each with the lowest penalty for its residual-risk level. The residual risk left by the selected method determines how much restriction is needed to reach a given risk level, while asset sensitivity, access rights, and operation criticality determine which restrictions achieve it at the lowest cost in functionality. Security administrators then select the configuration that matches their organization's risk appetite. We deliberately leave this decision to administrators: in domains such as healthcare and industrial control systems, restricting authorization can affect patient care and physical processes, so such changes should remain under accountable human control. Since the recommended configurations depend on context rather than the individual session, they can be computed in advance for each recurring context, such as emergency access from an ambulance, so a pre-approved configuration applies when that context occurs.

 We evaluate the approach using representative case studies from healthcare and industrial control systems (ICSs), compare it against common baseline authorization strategies, assess its robustness under plausible parameter uncertainty, and measure its computational overhead.
The results show that our approach can reduce both overall and attack-specific residual risks while preserving essential functionality. More importantly, rather than enforcing a single rigid authorization policy, the approach identifies multiple feasible authorization configurations with different risk--penalty trade-offs, thereby enabling security administrators to align authorization decisions with domain priorities and organizational risk appetite. This paper makes three contributions. First, it proposes an approach that extends adaptive authentication with adaptive authorization as a second line of defense, using the residual risk left by the selected authentication method to determine how far authorization must be restricted. Second, it presents a formal decision model that combines post-authentication residual risk, operation criticality, asset--operation dependencies, fine-grained access rights, and a penalty model distinguishing tightened access rights from disabled operations, together with an SMT-based procedure that characterizes the resulting risk--functionality trade-off as a set of recommended configurations. Third, it evaluates the approach on six healthcare and ICS scenarios, comparing it with three authorization baselines and analyzing its sensitivity to parameter uncertainty and its computational overhead.

The remainder of this paper is organized as follows. Section 2 introduces the motivating scenarios. Section 3 presents the proposed approach, including the adaptive authentication and adaptive authorization phases. Section 4 evaluates the approach using representative case studies. Section 5 discusses related work, and Section 6 concludes the paper and outlines directions for future research.

%% file: Scenarios.tex
\section{Motivating Scenarios in Healthcare}
\label{healthcare_scenarios}

\begin{figure*}[ht]
 	\centering
 	\begin{subfigure}[b]{0.33\linewidth}
 		\includegraphics[width=\linewidth]{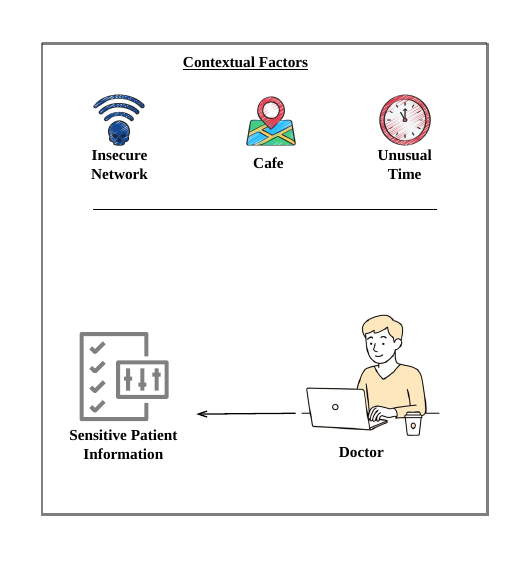}
 		\caption{Outside the Hospital (S1)}
 		\label{fig-10a}
 	\end{subfigure}
 	\begin{subfigure}[b]{0.33\linewidth}
 		\includegraphics[width=\linewidth]{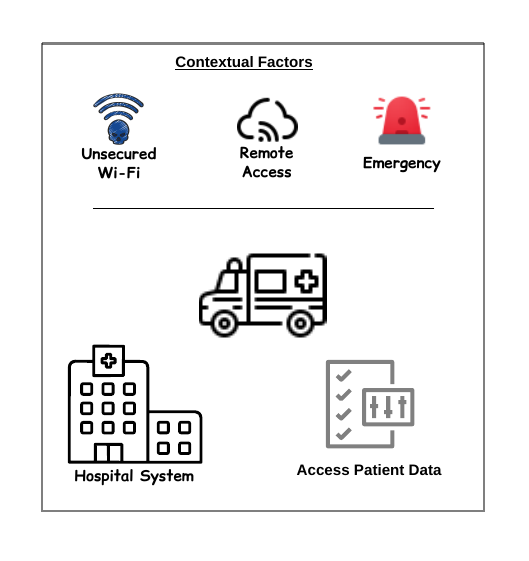}
 		\caption{Ambulance Emergency Access (S2)}
 		\label{fig-10b}
 	\end{subfigure}
 \begin{subfigure}[b]{0.33\linewidth}
 	\includegraphics[width=\linewidth]{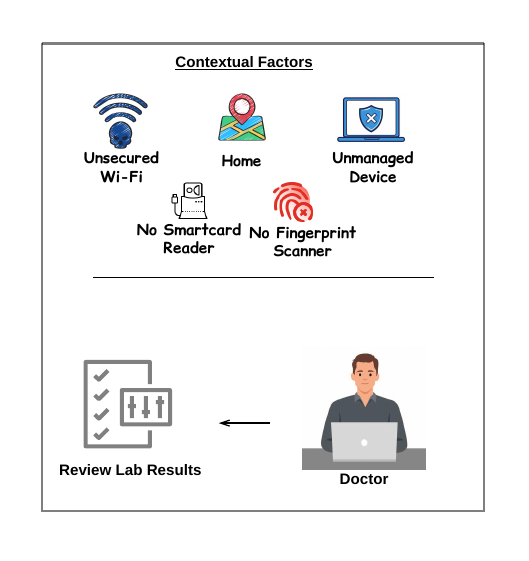}
 	\caption{Late Night Access (S3)}
 	\label{fig-10c}
 \end{subfigure}
 	\caption{Adaptive Authentication Scenarios in Healthcare.}
    \Description{Three panels showing the healthcare scenarios and their contextual factors. (a) Outside the hospital (S1): a doctor on a laptop accesses sensitive patient information over an insecure network in a cafe at an unusual time. (b) Ambulance emergency access (S2): patient data in the hospital system is accessed remotely from an ambulance over unsecured Wi-Fi during an emergency. (c) Late-night access (S3): a doctor reviews lab results from home on an unmanaged device over unsecured Wi-Fi, with no smartcard reader or fingerprint scanner available.}

 	\label{fig:healthcare_scenario}
 \end{figure*}

We consider the following healthcare scenarios (Figure~\ref{fig:healthcare_scenario}) to illustrate and motivate the need to account for residual risk in adaptive authentication. We choose the healthcare domain because it combines rich, dynamic contexts with highly sensitive data that attract attackers, making it ideal to motivate the residual risk problem.

\textbf{S1.}\label{S1} A family physician attempts to review lab results remotely from a caf\'e at an unusual time, over an insecure network; depending on the findings, the physician may also issue a prescription. This context suggests high risk, so the security requirement takes priority over usability and performance. The adaptive authentication system responds by requiring multi-factor authentication (MFA), such as a smartcard and a fingerprint.
 Although the second factor makes impersonation harder, fingerprint authentication remains vulnerable to the replay of captured biometric data, and the insecure network may still expose the session to hijacking~\cite{wang_attacks_2021,alaca_device_2016}. A successful attacker could obtain a session that can read and modify diagnoses, lab results, and prescriptions. This motivates restricting what an authenticated session can do, starting with lower-criticality operations such as adding notes and generating reports.

\textbf{S2.}\label{S2} A physician logs in to the system from an ambulance to retrieve patient information during a critical case. Despite the unfamiliar location and the use of an unsecured Wi-Fi network, the adaptive authentication system prioritizes usability and performance over security because of the emergency\footnote{A key consideration in this scenario is how context data, such as emergency situations, is made available to the adaptive authentication system. While acquiring context information is beyond this work’s scope, prior research shows that modern smart devices, like smartphones, are equipped with sensors capable of detecting environmental factors such as rapid movement, location changes, or abnormal access patterns \cite{arias-cabarcos_survey_2020}.} and therefore selects a single factor, for example, a smartcard. However, residual risk remains high. Over an unsecured Wi-Fi network, a smartcard alone leaves the session exposed to replay and session hijacking~\cite{fi15040146}, and a successful attacker could access patient records, diagnoses, and treatment plans. At the same time, restrictions must not impede care:  the physician must still be able to access the patient's profile, order tests, and adjust the care plan. This motivates restricting lower-priority operations, such as managing appointments or printing reports, while preserving those the emergency depends on.

\textbf{S3.}\label{S3} Late at night, a physician is asked to review a deteriorating patient's latest lab results and diagnosis remotely, using a personal, unmanaged laptop over unsecured home Wi-Fi, with no fingerprint scanner or smartcard reader available; depending on the findings, the physician may need to adjust the patient's medication. This context suggests high risk, so security requirements take priority over usability and performance. The adaptive authentication system selects a two-factor method supported by the available hardware: a hardware token plus a camera-based biometric, such as face or iris recognition. However, the selected method still leaves residual risk, mainly from replay attacks: camera-based biometrics can be deceived by replaying recorded images or video of the user, which malware on an unmanaged laptop could capture, and impersonation also remains possible~\cite {wang_attacks_2021,fi15040146}. This motivates restricting less critical operations and tightening access to sensitive assets after authentication.

In all three scenarios, adaptive authentication selects an appropriate method for the context, yet residual risk remains, and strengthening authentication further is either unacceptable, as in the emergency of S2, or not possible with the available hardware, as in S3. The response therefore has to act on what an authenticated session is allowed to do. The scenarios also show what such a response must achieve:
\begin{itemize}
    \item \textbf{R1 (Context-specific residual risk).} Different contexts lead to different methods, which leave different residual weaknesses: a smartcard and fingerprint used over a caf\'e's network (S1) are exposed to impersonation, replay, and session hijacking to different degrees than a smartcard alone in an ambulance (S2) or a token and a camera-based biometric on an unmanaged laptop (S3). Residual risk must therefore be quantified for the method selected in each context (Section~\ref{residual-risk}).
    \item \textbf{R2 (Preserve critical operations).} Some operations must remain available whatever the risk, such as accessing the patient's profile and adjusting the care plan in S2, while others, such as generating reports in S1 and S3, can be given up first. Operations must therefore be distinguished by their criticality (Section~\ref{sec:operation_classification}).
    \item \textbf{R3 (Fine-grained restriction).} The harm a compromised session can cause depends on which assets it can reach and with which access rights: a physician who only needs to read a diagnosis does not need the right to delete it. Restrictions should therefore apply to access rights for individual assets, not only to whole operations (Section~\ref{sec:asset-operation}).
    \item \textbf{R4 (Explicit trade-offs).} Every restriction removes some functionality, and how much functionality is worth giving up for a given reduction in residual risk differs between organizations. The response should therefore quantify the functionality lost and expose alternative configurations rather than impose a single one (Sections~\ref{sec:penalty} and~\ref{sec:decision-semantics}).
\end{itemize}

%% file: Approach.tex
\section{Approach for Mitigating Residual Risks in Adaptive Authentication }

Our approach extends adaptive authentication with adaptive authorization as a second line of defense to mitigate residual risks while preserving essential system functionality. It reduces the potential impact of authentication compromise by identifying authorization configurations that adapt permissions based on residual risk, operational criticality, and asset sensitivity, and uses an authorization penalty model to quantify the functionality loss of each adaptation.
Figure~\ref{fig:approach} illustrates the two main stages of our approach. First, the \textbf{adaptive authentication} phase selects the most appropriate authentication method based on contextual factors and system requirements \cite{10.1145/3828751}. Second, the \textbf{adaptive authorization} phase quantifies the residual risk left by the selected authentication method and recommends feasible authorization configurations that restrict access to operations and assets to reduce this risk (Section~\ref{sec:deployment}). We encode both phases in SMT-LIB and solve them using the Z3 SMT solver~\cite{10.1007/978-3-540-78800-3_24}. The adaptive authentication phase reuses the encoding from our previous work~\cite{10.1145/3828751}; the adaptive authorization phase introduces the encoding presented in this section.

\subsection{Threat Model and Assumptions}
\label{subsubsec:threatmodel}

We consider authentication-related threats in which an adversary attempts to access the system by bypassing or weakening authentication. We consider three classes of attacks: \textit{impersonation}, in which the adversary authenticates as a legitimate user by presenting stolen, guessed, or forged credentials or biometric traits; \textit{replay}, in which the adversary re-submits authentication data captured from a previous legitimate authentication; and \textit{session hijacking}, in which the adversary takes over a session that a legitimate user has already authenticated, for example by stealing its session token. We assume the adversary can interact with the authentication interface as an external user submitting login attempts and observing their outcomes, and may obtain credentials, authentication data, or session tokens through means such as leaked credentials, phishing, network exposure in unsecured environments, or a compromised user device. We do not model these means explicitly; their effect is reflected in the context-dependent attack likelihoods (Section~\ref{subsec:authn-phase}).

We assume that the system already enforces a baseline access-control policy, such as Role-Based Access Control (RBAC) or Attribute-Based Access Control (ABAC), that defines the operations a legitimate user can perform; we refer to it as the default authorization configuration $\pi_0$. Our adaptive authorization layer does not modify roles or attributes. Instead, it applies additional runtime restrictions on top of $\pi_0$ by disabling operations or tightening access rights, and we assume that these restrictions are enforced by a trusted enforcement point that the adversary cannot bypass.

Our primary attacker model is post-authentication compromise. After a successful authentication attack, the adversary holds a valid authenticated session and can invoke any operation permitted by the current authorization policy. Under this model, adaptive authorization reduces the potential \emph{impact} of a compromise, such as what can be modified or exfiltrated, by limiting access to sensitive assets and high-impact actions even if the session is compromised. Because the system cannot distinguish a compromised session from a legitimate one, any restriction applies to both: it limits what an attacker could do, but also what the legitimate user can do. This is why our approach quantifies the functionality lost by each restriction (Section~\ref{sec:penalty}) alongside the residual risk it removes.

We do not model threats that do not involve authentication, such as denial-of-service attacks, malware, misuse of administrator privileges by insiders, or attacks outside the application's authorization boundary, such as direct database compromise. Such threats require additional controls and are therefore outside the scope of this paper.

\subsection{Adaptive Authentication Phase}
\label{subsec:authn-phase}
\begin{figure*}[t]
    \centering    \includegraphics[width=1\linewidth]{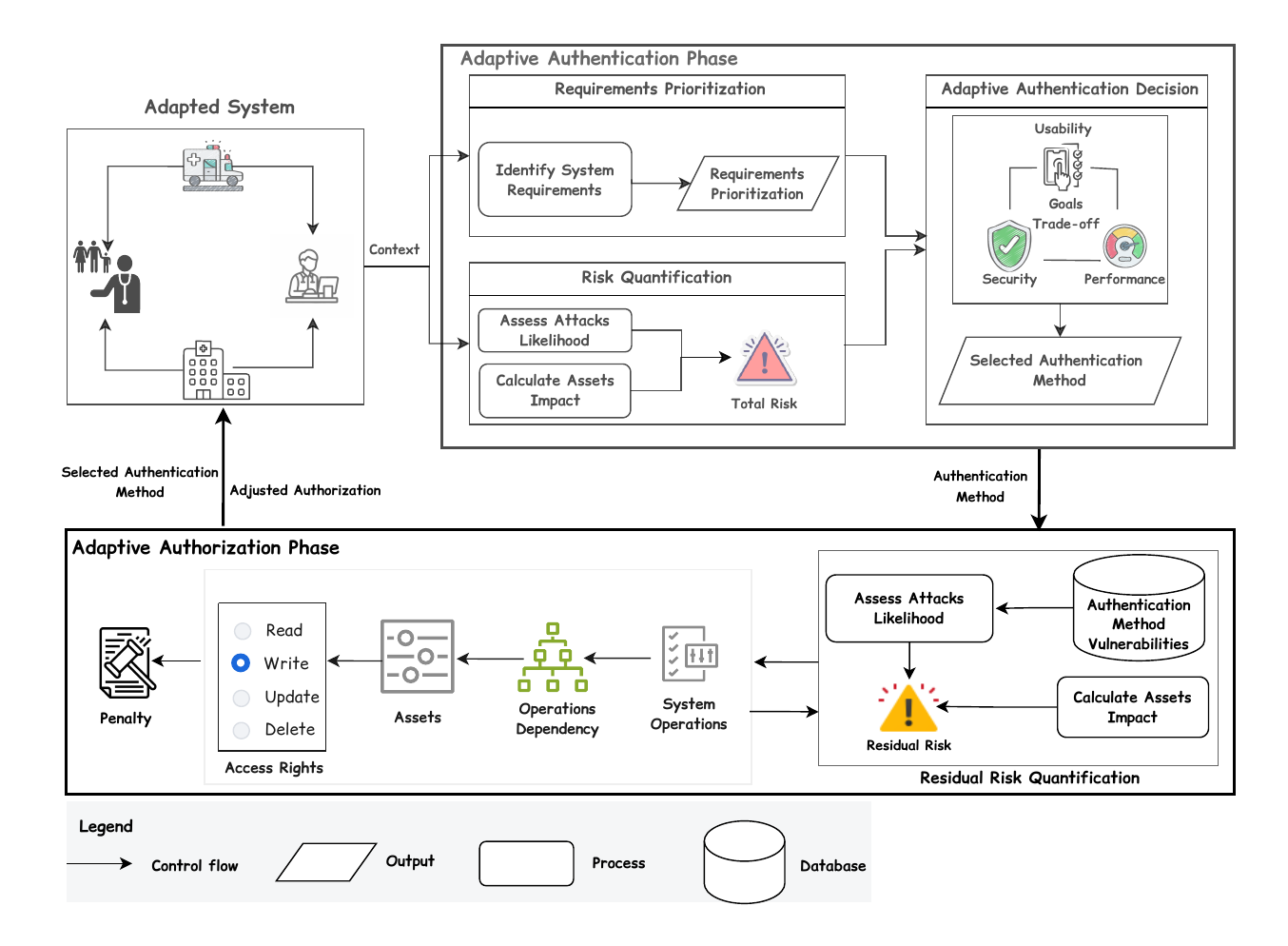}
    \caption{Adaptive authentication and authorization approach. The adaptive authentication phase (top) is adopted from our previous work~\cite{10.1145/3828751}; the adaptive authorization phase (bottom) is the extension proposed in this paper. The recommended configurations are presented to security administrators, who select or pre-approve the configuration to enforce (Section~\ref{sec:deployment}).}
    \Description{Diagram of the two-phase approach. The adaptive authentication phase uses the context to select an authentication method; the adaptive authorization phase quantifies the residual risk that the method leaves and derives authorization configurations together with their functionality penalties.}
    \label{fig:approach}
\end{figure*}

The first stage of our approach is the adaptive authentication phase (Figure~\ref{fig:approach}). This phase builds directly on our previous requirement-based adaptive authentication framework~\cite{10.1145/3828751}. In that framework, a contextual goal model represents authentication requirements and the impact of contextual factors on security risks, requirement priorities, and the feasibility of authentication methods. An extended feature model represents available authentication methods and their impact on requirement satisfaction. We use a Fuzzy Causal Network (FCN) encoded in Z3 to reason about these relations and select an authentication method appropriate to the current context. The framework is a suitable foundation for the present work, since it already provides context-aware authentication-method selection and the information needed to quantify the residual risk remaining after authentication.

Contextual factors influence the priority of authentication requirements in the contextual goal model. For example, connecting from an unknown network or location increases security risks and raises the priority of security requirements, whereas accessing data in an emergency situation increases the priority of performance and usability requirements. For a leaf requirement goal $g$, its priority $\rho(g)$ is computed from the contextual factors associated with it as:

\begin{equation}
\rho(g) = \max_{1 \le i \le N} \operatorname{impact}(c_i, g),
\quad CF(g)=\{c_1,\dots,c_N\},
\label{eq:goal-priority}
\end{equation}

where $CF(g)$ is the set of contextual factors associated with goal $g$, and $\operatorname{impact}(c_i, g)$ denotes the contribution of contextual factor $c_i$ to the priority of $g$. Thus, contextual changes dynamically affect which requirements become more important when selecting an authentication method.

Contextual factors influence the likelihood of authentication-related attacks before an authentication method is applied. For a modeled attack $a_i$, this likelihood is derived from the contextual factors associated with $a_i$. If more than one contextual factor affects the same attack, their contributions are conservatively aggregated using the maximum operator. Contextual factors can also influence the harm caused by a successful attack, which our model captures through asset sensitivities and weights that may differ between scenarios (Section~\ref{residual-risk}). Accordingly, before authentication  is applied, we define the total risk of an attack $a_i$ as:
\begin{equation}
R^{a_i}_{\text{before}} = L^{a_i}_{\text{before}} \cdot I(\pi_0),
\label{eq:total-risk}
\end{equation}

where $L^{a_i}_{\text{before}} \in [0,1]$ is the likelihood of $a_i$ under the current contextual conditions, and $I(\pi_0) \in [0,1]$ is the impact of a successful attack under the default authorization configuration $\pi_0$, as defined in Eq.~\ref{eq:impact}. The default authorization configuration is the access-control policy that the system already enforces (Section~\ref{subsubsec:threatmodel}), before any adaptive authorization restriction is applied. It is the unrestricted member of the set of authorization configurations considered in this paper: every operation is enabled and every asset retains its original access rights, so it incurs no functionality loss. In our case studies, $\pi_0$ grants all four access rights on every asset, which is the most permissive starting point. Where the baseline policy already withholds some rights, for example, Delete on clinical records, the original access rights of each asset are set accordingly, and the approach only restricts rights that the baseline actually grants.

Based on these context-dependent priorities and changes, the adaptive authentication framework evaluates how the available authentication methods satisfy the system requirements and selects the method that maximizes the satisfaction of the requirements and minimizes security risks. For further details on how the optimal authentication method is selected in the adaptive authentication phase, we refer the reader to our previous work~\cite{10.1145/3828751}.

Although this phase selects an authentication method appropriate to the current context, it does not address the residual risk that may remain after authentication. We therefore use this phase as the first stage of the overall approach and reuse two of its outputs for each modeled attack $a_i$: (i) the context-dependent pre-mitigation likelihood $L^{a_i}_{\text{before}}$, and (ii) the selected authentication method, whose mitigation effectiveness against $a_i$ determines the post-authentication likelihood $L^{a_i}_{\text{after}}$ (Section~\ref{residual-risk}). Any adaptive or risk-based authentication mechanism that provides them could therefore replace the framework~\cite{10.1145/3828751} that we use as one instantiation.
\subsection{Adaptive Authorization Phase}
In this phase, the approach identifies authorization configurations that reduce the residual risk left by the selected authentication method. First, it quantifies this residual risk for the selected method (Section~\ref{residual-risk}; R1). Next, it classifies system operations as \textit{Must-Have}, \textit{Nice-to-Have}, or \textit{Optional} (Section~\ref{sec:operation_classification}; R2) and links them to the assets and access rights required for their execution (Section~\ref{sec:asset-operation}; R3). Finally, it quantifies the functionality lost through authorization restrictions (Section~\ref{sec:penalty}) and searches for configurations that reduce residual risk at the lowest cost in functionality, which it presents to security administrators as a set of recommended configurations (Sections~\ref{sec:decision-semantics} and~\ref{sec:deployment}; R4).

\subsubsection{Residual Risk Quantification}
\label{residual-risk}

To formalize the notion of residual risk in our approach, we draw upon definitions established in information security and risk management research~\cite{9590416,app13010395}. In this work, residual risk is defined as the portion of risk that remains after applying security controls. We adapt this definition to the context of adaptive authentication, where each authentication method serves as a security control that mitigates security risks.
For each authentication-related attack $a_i$ (e.g., impersonation, replay, session hijacking), we distinguish two quantities: the total risk \emph{before} an authentication method is applied, computed in the adaptive authentication phase (Eq.~\ref{eq:total-risk}), and the residual risk that remains \emph{after} the authentication method is applied. Once adaptive authentication selects an authentication method, the likelihood of $a_i$ is reduced to $L^{a_i}_{\text{after}}$, and the residual risk under an authorization configuration $\pi$ is

\begin{equation}
R^{a_i}_{\text{residual}}(\pi) = L^{a_i}_{\text{after}} \cdot I(\pi),
\label{eq:residual-risk}
\end{equation}

where $L^{a_i}_{\text{after}} \in [0,1]$ is the likelihood of attack $a_i$ after the selected authentication method is applied, and $I(\pi) \in [0,1]$ is the impact of a successful attack under $\pi$.

For the authentication attacks considered in this work, we use the same authorization impact $I(\pi)$ across attack types. This follows from our threat model in Section~\ref{subsubsec:threatmodel}: a successful impersonation, replay, or session-hijacking attack results in an authenticated session whose subsequent capabilities are determined by the same authorization configuration $\pi$. Therefore, these attacks differ in their likelihoods and in the effectiveness of the selected authentication method against them, but, once successful, they expose the attacker to the same set of authorized operations, assets, and access rights. Consequently, their authorization impact is modeled identically. %

In our implementation, $L^{a_i}_{\text{before}}$ is a context-dependent likelihood derived from modeled contextual risk factors (e.g., unsecured Wi-Fi or an unfamiliar location increases the baseline likelihood for replay/session hijacking). As in our previous work~\cite{10.1145/3828751}, the effect of each contextual factor on an attack, and the effectiveness of each authentication method in mitigating it, are first assessed qualitatively from the literature on authentication attacks and contextual impacts (e.g., \cite{alaca_device_2016, Wiefling_Pump_2022,bumiller_towards_2022}) and from expert judgment, and then mapped to a value in $[0,1]$: values above $0.8$ denote a very strong effect ($++$), values in $(0.6, 0.8]$ a strong effect ($+$), values in $(0.4, 0.6]$ a moderate effect, values in $(0.2, 0.4]$ a weak effect ($-$), and values of $0.2$ or below a very weak effect ($--$). The post-authentication likelihood $L^{a_i}_{\text{after}}$ is obtained by subtracting the mitigation effectiveness of the selected authentication method against $a_i$ from $L^{a_i}_{\text{before}}$, bounded below by zero. Mitigation therefore reduces the likelihood of an attack but does not generally eliminate it; the remaining likelihood is what gives rise to the residual risk that adaptive authorization mitigates. For a two-factor method combining $m_1$ and $m_2$, this effectiveness is $\min\{1, (c_{a_i,m_1} + c_{a_i,m_2} + b)/2\}$, where $c_{a_i,m}$ is the mitigation coefficient of method $m$ against $a_i$ and $b = 0.30$ is a two-factor bonus. When precise empirical values are unavailable, we use bounded parameter ranges ($[0,1]$) and choose conservative assignments that avoid underestimating risk. We then assess stability via sensitivity analysis by perturbing key model parameters (Section~\ref{sec:healthcare_sensitivity}). Because these values are obtained by mapping qualitative ratings onto $[0,1]$, the resulting residual-risk values should be read as a relative index for comparing authorization configurations within a scenario, rather than as calibrated probabilities of compromise. Accordingly, the evaluation reports residual-risk reductions relative to the default configuration $\pi_0$.

\paragraph{Impact as authorization exposure.}
In this paper, an asset is any protected system resource of value, such as patient records, diagnoses, or system logs, that a user operation may access or affect. We model impact as dependent on the authorization configuration because our mitigation acts primarily on post-compromise capabilities. Intuitively, a compromised session is more harmful if it retains powerful actions (e.g., \textit{Update} or \textit{Delete}) over highly sensitive assets. Therefore, $I(\pi)$ is computed from both the assets that remain accessible under configuration $\pi$ and the access rights granted on those assets. Specifically, each asset is assigned a sensitivity value based on the potential consequences of unauthorized disclosure, modification, or misuse. For example, clinical decision assets such as Diagnosis are assigned a high sensitivity ($val = 0.90$) because they directly affect care, whereas administrative assets such as ReferralNotes are assigned a lower sensitivity ($val = 0.68$). At the same time, access rights affect exposure; for example, retaining \textit{Update} on an asset gives an exposure of $0.75$, compared with $0.25$ for \textit{Read-only} access, as defined below.

To compute the authorization exposure impact, let $S$ denote the set of protected assets in the system, and let $S_\pi \subseteq S$ denote the assets that remain enabled under authorization configuration $\pi$. For each asset $s$, let $val(s)\in[0,1]$ denote its sensitivity, representing the potential consequence of its unauthorized access. Let $exp(s,\pi)\in[0,1]$ denote the exposure associated with the access rights retained on asset $s$ under configuration $\pi$.

\begin{equation}
I(\pi)=\sum_{s \in S_\pi}
w_s \cdot val(s) \cdot exp(s,\pi),
\label{eq:impact}
\end{equation}

where $w_s \geq 0$ represents the relative contribution of asset $s$ to the overall impact, with $\sum_{s\in S} w_s = 1$. The asset weights are defined over the complete set of assets. %
Consequently, disabling an asset removes its contribution to $I(\pi)$.

Exposure is determined by the strongest access right retained on the asset. For the instantiation used in this paper, we set $exp(s,\pi) = 0.25$, $0.50$, $0.75$ and $1.00$ for Read, Write, Update and Delete respectively, and $exp(s,\pi) = 0$ when the asset is disabled. This is a uniform ordinal scale over the access-right hierarchy of Section~\ref{sec:asset-operation}, so that each additional level of access contributes an equal increment of exposure. Since a higher access right implies the lower ones, the strongest retained right determines exposure, consistent with the access-rights tightness in Eq.~\ref{eq:permission-tightness}. This mapping is an instantiation rather than a general assumption; domains with different access-right structures may define a different exposure scale. Consequently, $I(\pi)$ is higher when more sensitive assets remain accessible and when stronger access rights are retained on those assets. This scale ranks access rights by the damage they allow to an asset's integrity. For confidentiality, even Read access to a highly sensitive asset can be harmful; our model reflects this partly through the sensitivity $val(s)$, and domains in which disclosure is the main concern can assign a higher exposure to Read.

Asset sensitivities and impact weights are provided as inputs to the model. Sensitivity $val(s)$ reflects the consequence of unauthorized access to the asset in the application domain. The weight $w_s$ reflects the asset's relative contribution to overall impact, judged by how central it is to the Must-Have operations of the scenarios in which it appears; weights may therefore differ across scenarios.
For example, in S1, diagnostic and treatment-critical assets carry the largest weights (Diagnosis 0.15, TreatmentPlan 0.13), whereas administrative assets such as ReferralNotes and EncounterNotes carry the smallest (0.05 each). Both quantities are assigned through expert judgment informed by the domain rather than derived from an empirical procedure. We treat them as modeling inputs and assess the influence of the sensitivities through the sensitivity analysis in Section~\ref{sec:healthcare_sensitivity}.

The residual risk is computed as likelihood $\times$ impact, a standard risk formulation in security risk assessment \cite{nist_sp800_30r1}. This captures both the effect of security controls on attack likelihoods and the potential harm to assets in the event of a compromise. Each authentication method reduces the likelihood of specific attacks, depending on its effectiveness against those attacks and the context in which it is applied.

Because impact depends on authorization, residual risk is evaluated at two points in our approach: first, under $\pi_0$, to quantify the exposure left by the authentication before authorization adaptation, and then under each candidate configuration $\pi$ during the adaptive authorization phase. Since authorization adaptation changes the impact term $I(\pi)$, different configurations yield different residual-risk values and expose different risk–penalty trade-offs.

We conservatively aggregate the attack-specific residual risks using the maximum operator:

\begin{equation}
R_{\mathrm{residual}}(\pi)
=
\max_{a_i \in A}
R^{a_i}_{\mathrm{residual}}(\pi).
\end{equation}

This aggregation adopts a worst-case criterion in which the attack with the highest remaining residual risk determines the overall residual risk. We acknowledge that max-based aggregation may lead to stricter authorization decisions. This choice of Max follows from our threat model (Section~\ref{subsubsec:threatmodel}): because every successful attack yields an authenticated session with the same capabilities, an adversary needs only one attack to succeed, and averaging would understate the exposure left by the most likely one.
At the same time, to illustrate how our approach behaves under a different aggregation function, we also consider the average aggregation strategy in the ICS case study. This shows that the approach is not tied to a single aggregation function and that its outputs vary in interpretable ways depending on how per-attack residual risks are combined. More generally, alternative aggregation strategies, such as averages or weighted sums, could be adopted depending on risk aversion. Our model is modular with respect to risk aggregation: replacing the max operator with an alternative aggregation function does not require structural changes to the model or its SMT encoding; only the risk aggregation function needs to be changed.

\begin{table*}[b]
\centering
\caption{Operation Classification \& Asset-Operation Dependencies in the healthcare scenarios}
\footnotesize
\renewcommand{\arraystretch}{1.13}

\begin{tabular}{|p{3.2cm}|>{\centering\arraybackslash}p{1.5cm}|>{\centering\arraybackslash}p{1.7cm}|>{\centering\arraybackslash}p{1.4cm}|p{5.5cm}|}
\hline
\textbf{Operation} & \textbf{S1: Outside the hospital} & \textbf{S2: Ambulance Emergency Access} & \textbf{S3: Late Night Access} & \textbf{Associated Assets} \\
\hline

AccessPatientProfile & -- & M & -- &
PatientRecords, PatientDemographics, PatientHistory \\

\hline
OrderLabTest & -- & M & -- &
LabResults, TestReports, PatientRecords \\

\hline
AdjustCarePlan & -- & M & -- &
TreatmentPlan, Diagnosis, PatientRecords \\

\hline
DiagnoseMedicalConditions & M & -- & M &
Diagnosis, TreatmentPlan \\

\hline
ViewLabResults & M & -- & M &
LabResults, TestReports \\

\hline
ApproveDischargeSummary & -- & N & -- &
DischargeSummaries, TreatmentPlan \\

\hline
RequestConsultation & -- & N & -- &
ReferralNotes, EncounterNotes, PatientRecords \\

\hline
CommunicateWithPatient & -- & N & -- &
PatientCommunications, PatientInteractions \\

\hline
PrescribeControlled & N & -- & N &
ControlledPrescriptions, PatientHistory, Diagnosis \\

\hline
PrescribeNonControlled & N & -- & N &
NonControlledPrescriptions, PatientHistory, Diagnosis \\

\hline
ViewRadiologyImages & -- & O & -- &
RadiologyImages, TestReports, PatientRecords \\

\hline
ManageAppointment & -- & O & -- &
VisitSummaries, PatientRecords \\

\hline
PrintMedicalReport & -- & O & -- &
Diagnosis, TreatmentPlan, LabResults, PatientRecords \\

\hline
AddEncounterNote & O & -- & O &
EncounterNotes, PatientInteractions, VisitSummaries \\

\hline
AddReferralNote & O & -- & O &
ReferralNotes, EncounterNotes \\

\hline
GenerateReports & O & -- & O &
Diagnosis, TreatmentPlan, LabResults, PatientRecords \\

\hline
\end{tabular}
\footnotesize \textbf{*Note:} M = Must-Have, N = Nice-to-Have, O = Optional. Under $\pi_0$, every listed asset carries all four access rights (R, W, U, D).
\label{tab:healthcare_operation_criticality_assets_all}
\end{table*}

\subsubsection{Operation Classification}
\label{sec:operation_classification}
To systematically decide which operations to restrict to reduce residual risk, we classify system operations into three classes: \textit{Must-Have}, \textit{Nice-to-Have}, and \textit{Optional}. This classification (inspired by common software requirement prioritization schemes \cite{Berander2005}) reflects the criticality of each operation to the system’s mission. \textit{Must-Have} operations are those critical to core functionality. \textit{Nice-to-Have} operations are important but not mission-critical. \textit{Optional} operations are low-priority features. This classification guides the adaptive authorization phase by indicating which operations are considered first for restriction to reduce exposure while preserving system functionality.

\paragraph{Classification procedure.} To classify operations, we follow a lightweight, traceable procedure per scenario: (i) identify the scenario goal and its operations, (ii) determine whether disabling each operation prevents achieving the goal, and (iii) label operations as \textit{Must-Have} if goal completion would be blocked, \textit{Nice-to-Have} if goal completion can still be achieved partially, and \textit{Optional} if disabling the operation does not affect goal completion. This procedure provides an explicit rationale for each assigned label, and Table~\ref{tab:healthcare_operation_criticality_assets_all} and Table~B.1 of the supplementary material present the resulting classifications and asset dependencies for the healthcare and ICS scenarios, respectively.
For example, in S1 the goal is to review a patient's lab results and update the diagnosis. Disabling \textit{DiagnoseMedicalConditions} or \textit{ViewLabResults} would block this goal, so both are Must-Have. Disabling a prescription operation still allows the review to be completed but prevents its possible follow-up, so both prescription operations are Nice-to-Have. Disabling \textit{AddEncounterNote}, \textit{AddReferralNote}, or \textit{GenerateReports} does not affect the goal, so these operations are Optional.

The same operation may receive different labels across scenarios depending on the scenario's goal and contextual constraints. These labels are assigned prior to solving and are used as inputs to guide the adaptive authorization phase. Historical access logs may also support this process; we leave automated extraction of such labels to future work.
Only Must-Have operations are protected by a hard constraint: they are never disabled (constraint (C1) in Section~\ref{sec:decision-semantics}). An operation's class determines its penalty weight $w(o)$ (Section~\ref{sec:penalty}), and because Optional operations carry the smallest weight, disabling one of them is usually the cheapest way to reduce residual risk further. Optional operations are therefore typically disabled before Nice-to-Have ones.
This graded approach manages residual risk by sacrificing less-critical operations before restricting the most critical ones.

\subsubsection{Asset-Operation Dependencies and Access Rights on Assets}
\label{sec:asset-operation}

In this step, our approach maps each operation to the assets it uses and the access rights it requires on them \cite{kizza2024access}. This reduces exposure by disabling an operation or by tightening access rights on its assets while the operation remains enabled. We model authorization at both levels because the impact of a compromise depends on which access rights remain available to the protected assets.

The relationships among access rights depend on the application domain. They are provided as input to the approach and can therefore be configured to meet the requirements of the application domain.
For the purposes of the healthcare and ICS case studies considered in this paper, we use the access rights \textit{Read}, \textit{Write}, \textit{Update}, and \textit{Delete}, and model them using the hierarchical relationship \textit{Delete} $\rightarrow$ \textit{Update} $\rightarrow$ \textit{Write} $\rightarrow$ \textit{Read}. Here, \textit{Write} denotes adding new entries to an asset (e.g., adding an encounter note) and therefore corresponds to the \textit{Create} operation of the CRUD model, whereas \textit{Update} denotes modifying existing entries. Under this instantiation, granting a higher-level access right implies that the dependent lower-level access rights are also granted, and revoking a lower-level access right automatically revokes any dependent higher-level access rights. This prevents inconsistent authorization states within this access rights structure, such as enabling \textit{Update} without \textit{Read}. For example, retaining \textit{Update} on TreatmentPlan also retains \textit{Write} and \textit{Read} on it, while revoking \textit{Read} on TreatmentPlan also revokes \textit{Write}, \textit{Update}, and \textit{Delete}. Consequently, the access rights retained on any asset always form one of four nested levels: (R), (R,W), (R,W,U), or (R,W,U,D), which is how they are reported in the configuration tables of the evaluation. This fine-grained representation is necessary because restricting an operation and tightening its underlying assets' access rights can affect residual risk and functionality differently.

Assets are considered active only if they are required by at least one enabled operation. When an asset is inactive, all its access rights are disabled. When it is active, the access rights required by at least one enabled operation that depends on that asset must remain available. This relationship directly links asset exposure to the system's operational needs, as shown in Table~\ref{tab:healthcare_operation_criticality_assets_all}. Each operation specifies the minimum set of assets and their access rights to function. For example, in the healthcare case study, accessing a patient profile requires \textit{Read} access to the corresponding PatientHistory asset. This mapping enables our approach to reason about the exposure associated with each operation based on both asset sensitivity and access rights.
By representing access rights as Boolean variables, the model can reason about which operations or access rights may be restricted to mitigate residual risk while preserving the most critical operations. These mappings and domain-specific access rights relationships are encoded as structural constraints in the SMT encoding (constraints (C2)--(C4) in Section~\ref{sec:decision-semantics}), ensuring that candidate authorization configurations respect operation--asset and access rights dependencies. This structure also provides the basis for computing authorization penalties associated with reduced access rights or disabled operations, as discussed next.

\subsubsection{Penalty-Based Model}
\label{sec:penalty}
Residual risk mitigation through adaptive authorization may reduce system functionality by either restricting access rights granted on assets or disabling operations entirely. We therefore define an authorization penalty that quantifies the functionality loss introduced by an authorization configuration $\pi$. The penalty distinguishes between these two forms of restriction while ensuring that disabling an operation is always considered more disruptive than restricting the access rights available to the assets associated with that operation.

We first quantify the degree to which access rights on an asset are restricted, which we call access-rights tightness and define in Eq.~\ref{eq:permission-tightness}. Let $\ell_0(s)$ denote the highest access-right level available for asset $s$ before adaptive authorization is applied, and let $\ell_{\pi}(s)$ denote the highest access-right level retained under configuration $\pi$.
Our approach assumes that the relevant access rights structure and the corresponding restriction levels are specified as part of the domain model. In the healthcare and ICS case studies considered in this paper, we use the access rights Read, Write, Update, and Delete and model them using the hierarchical ordering \(Read < Write < Update < Delete\). This ordering is an instantiation of the model rather than a general assumption; other domains may define different access rights relationships. For example, an audit system may permit append/write access without granting read or update access.
For this instantiation, we associate the access-right levels with the ranks
\[
    r(\text{Read})=0,\qquad
    r(\text{Write})=1,\qquad
    r(\text{Update})=2,\qquad
    r(\text{Delete})=3.
\]
For an asset that remains active under $\pi$, we define its access-rights tightness as
\begin{equation}
\operatorname{tight}(s,\pi)=
\begin{cases}
\dfrac{
r(\ell_0(s))-r(\ell_{\pi}(s))
}{
r(\ell_0(s))
},
& \text{if } r(\ell_0(s))>0, \\[8pt]
0,
& \text{if } r(\ell_0(s))=0 .
\end{cases}
\label{eq:permission-tightness}
\end{equation}
Since adaptive authorization only restricts existing access rights and does not grant rights beyond those originally available, we require \(r(\ell_{\pi}(s))\leq r(\ell_0(s))\). Therefore, $\operatorname{tight}(s,\pi)\in[0,1]$, where $0$ indicates that the original access rights are preserved and $1$ indicates the maximum possible tightening relative to the access rights originally available. For example, if an asset initially allows Read, Write, Update, and Delete, retaining only Read results in $\operatorname{tight}(s,\pi)=1$, whereas retaining Read, Write, and Update results in $\operatorname{tight}(s,\pi)=1/3$. If the asset is Read-only before adaptation, retaining Read incurs no access rights penalty.

We next quantify access rights restriction at the operation level. Let $\mathcal{O}$ denote the set of operations and let $S_o$ denote the set of assets associated with operation $o$. For an enabled operation, its access-restriction score is defined as the average tightness of the assets on which it depends:

\begin{equation}
T_o(\pi)
=
\frac{1}{|S_o|}
\sum_{s\in S_o}
\operatorname{tight}(s,\pi),
\qquad
T_o(\pi)\in[0,1].
\label{eq:operation-permission-loss}
\end{equation}

This formulation captures the degree to which access rights to the assets associated with an operation are restricted while the operation remains available. Since an enabled operation always retains the minimum access rights it requires (Section~\ref{sec:asset-operation}), tightening the access rights on its assets does not block the operation but reduces what it can do with them, and this reduction grows with both how many of its assets are tightened and how far. We therefore use the average tightness, which reflects both and keeps $T_o(\pi)$ within $[0,1]$ regardless of how many assets the operation uses. Assets associated with disabled operations are not separately penalized through \(T_o(\pi)\), since the loss of the corresponding operation is accounted for directly by the operation-disablement term in the overall penalty below.

Let $e_o(\pi)\in\{0,1\}$ indicate whether operation $o$ is enabled under configuration $\pi$, where $e_o(\pi)=1$ denotes an enabled operation. Each operation is assigned a criticality weight $w(o)>0$, with larger weights assigned to Must-Have operations than to Nice-to-Have ones, which in turn receive larger weights than Optional operations. For example, in our case studies, we use $w(o) = \mathbf{0.9}$ for Must-Have, $\mathbf{0.6}$ for Nice-to-Have, and $\mathbf{0.3}$ for Optional. The overall authorization penalty is then defined as

\begin{equation}
P(\pi)
=
\frac{
\displaystyle
\sum_{o\in\mathcal{O}}
w(o)
\left[
(1-e_o(\pi))
+
\beta\,e_o(\pi)T_o(\pi)
\right]
}{
\displaystyle
\sum_{o\in\mathcal{O}}w(o)
}, \qquad 0\leq\beta<1.
\label{eq:authorization-penalty}
\end{equation}

The first term, $1-e_o(\pi)$, captures the functionality loss caused by disabling an operation. The second term, $\beta e_o(\pi)T_o(\pi)$, captures the loss caused by tightening access rights while the operation remains enabled. The parameter $\beta$ controls the relative contribution of access rights tightening. Since $\beta<1$ and $T_o(\pi)\leq1$, disabling an operation always incurs a larger penalty for that operation than any access rights tightening that allows it to remain enabled.

The resulting penalty is normalized such that $P(\pi)\in[0,1]$. A configuration in which every operation is enabled and every asset retains its original access rights has $P(\pi)=0$, whereas disabling all operations results in $P(\pi)=1$. Furthermore, the penalty is monotonic at the operation level: tightening access rights associated with an enabled operation cannot decrease that operation's contribution to $P(\pi)$, while disabling an enabled operation incurs the maximum penalty contribution for that operation.

For an enabled operation, the contribution of access-rights tightening is bounded by $\beta w(o)$, whereas disabling the operation contributes $w(o)$. Thus, smaller values of $\beta$ make access rights tightening on assets comparatively inexpensive, while values closer to $1$ treat severe restrictions as approaching the functionality loss of disabling the operation. We treat $\beta$ as a modeling parameter rather than an empirically established constant; we use $\beta = 0.75$ in the case studies and its influence on the results is examined in Section~\ref{sec:healthcare_sensitivity}.

Unlike the residual risk quantification, the penalty does not use asset sensitivity. Asset sensitivity represents the potential security impact of compromise and is therefore incorporated into residual-risk quantification, whereas the authorization penalty captures the system functionality loss. Separating these quantities allows the approach to characterize the trade-off between reducing security risk and preserving operational capability without conflating security impact with functional importance.

\subsubsection{Decision Semantics and Feasible Configuration Space}
\label{sec:decision-semantics}

The adaptive authorization layer does not aim to produce a single ``optimal'' policy. Instead, it identifies a set of \emph{feasible} authorization configurations that satisfy security-related constraints, operational dependencies, and authorization-consistency rules. Rather than defining a single global objective that jointly optimizes residual risk and functionality loss, our approach provides a set of feasible configurations in terms of two descriptive metrics: (i) the residual risk remaining after applying authorization adjustments, and (ii) the authorization penalty induced by restricting access rights or disabling operations. These metrics are used to characterize the feasible space and expose the risk–penalty trade-offs among candidate configurations. We adopt this design because residual risk and functionality loss have no common scale. Combining them into a single objective would require fixing how much functionality a given reduction in residual risk is worth, and this trade-off depends on each organization's risk appetite and operational constraints. Presenting a set of configurations therefore leaves this judgment to security administrators rather than embedding an arbitrary weighting in the approach.

Formally, an authorization configuration $\pi$ assigns to each operation $o \in \mathcal{O}$ an enablement value $e_o(\pi) \in \{0,1\}$, and to each asset $s \in S$ a retained access level $\ell_\pi(s) \in \{\bot, \textit{Read}, \textit{Write}, \textit{Update}, \textit{Delete}\}$, where $\bot$ denotes a disabled asset and we extend the ranks of Section~\ref{sec:penalty} with $r(\bot) = -1$. Let $\mathcal{O}_M \subseteq \mathcal{O}$ denote the Must-Have operations and $\mathit{req}(o,s)$ the minimum access level that operation $o$ requires on asset $s \in S_o$. A configuration is admissible if it satisfies
\begin{align}
& e_o(\pi) = 1 && \forall o \in \mathcal{O}_M, \tag{C1}\\
& e_o(\pi) = 1 \;\Rightarrow\; r(\ell_\pi(s)) \ge r(\mathit{req}(o,s)) && \forall o \in \mathcal{O},\ \forall s \in S_o, \tag{C2}\\
& \ell_\pi(s) = \bot \;\Leftrightarrow\; \big(\forall o \in \mathcal{O} : s \in S_o \Rightarrow e_o(\pi) = 0\big) && \forall s \in S, \tag{C3}\\
& r(\ell_\pi(s)) \le r(\ell_0(s)) && \forall s \in S. \tag{C4}
\end{align}
Constraint (C1) preserves the Must-Have operations, (C2) ensures that every enabled operation retains the access rights it requires, (C3) disables exactly the assets that no enabled operation uses, and (C4) prevents any access right from being granted beyond the default configuration. Together with the access-right hierarchy of Section~\ref{sec:asset-operation}, these constraints are the operational and authorization-consistency constraints referred to above.

To generate candidate configurations, we progressively reduce the target residual risk and search for feasible authorization configurations within successive residual-risk intervals. Let $R_0 = R_{\mathrm{residual}}(\pi_0)$ denote the residual risk left after adaptive authentication, and let $\Delta>0$ denote the residual-risk reduction step. For each search step $k$, we define the lower boundary of the residual-risk interval as

\begin{equation}
\tau_k = \max(0, R_0-k\Delta).
\end{equation}

The corresponding search interval is therefore $[\tau_k,\tau_{k-1})$, where $\tau_0=R_0$. A configuration $\pi$ is considered admissible for interval $k$ if its residual risk falls within this interval and it satisfies the operational and authorization-consistency constraints. Among the admissible configurations within the same residual-risk interval, we identify the configuration with the lowest authorization penalty:

\begin{equation}
\pi^{*}_k
=
\arg\min_{\pi} P(\pi)
\quad
\text{s.t. }
\tau_k
\leq
R_{\mathrm{residual}}(\pi)
<
\tau_{k-1}.
\label{eq:arg_min}
\end{equation}

To solve Eq.~\ref{eq:arg_min} for each residual-risk interval $[\tau_k,\tau_{k-1})$, we encode the interval boundaries as SMT constraints and invoke Z3's optimization engine to minimize the authorization penalty $P(\pi)$ over the feasible authorization configurations.
The search process first considers authorization configurations that reduce the residual risk from $R_0$ to a value in the interval $[R_0-\Delta,R_0)$; the next search step considers the interval $[R_0-2\Delta,R_0-\Delta)$; and so forth.
For each interval containing one or more admissible configurations, Z3 returns a configuration attaining the minimum value of the authorization penalty $P(\pi)$ for that interval; when multiple configurations attain that minimum, Z3 returns one of them, which we retain as the recommended configuration for the interval. This ensures that configurations providing comparable levels of residual-risk reduction are distinguished by the functionality loss they introduce, rather than by the order in which Z3 returns them. This interval-based search is closely related to the $\varepsilon$-constraint method for bi-objective optimization~\cite{ehrgott_improved_2008}, which minimizes one objective while bounding the other. Because each interval is solved independently, the search does not by itself guarantee that the recommended configurations are mutually non-dominated. However, in all six evaluation scenarios, the authorization penalty increases monotonically as the residual risk decreases, so that all recommended configurations are mutually non-dominated.

Consequently, the number of reported configurations is not fixed. It depends on the residual risk left after adaptive authentication, the reduction step $\Delta$, and the feasible authorization configurations available in each scenario. Enumeration continues toward lower residual-risk intervals until the lower bound reaches zero or Z3 establishes that no feasible authorization configuration exists with a residual risk below the current search range.

In this paper, we refer to the lowest-penalty admissible configuration identified within a given residual-risk interval as the \emph{recommended configuration}. The approach, therefore, does not prescribe a single globally optimal authorization policy. Instead, it exposes a set of recommended configurations associated with progressively stronger residual-risk reductions, allowing security administrators to select a configuration according to their acceptable residual-risk level and functionality constraints.

\subsubsection{Deployment Model and Decision Support}
\label{sec:deployment}

The proposed approach is intended as a decision-support mechanism for security administrators, not a fully autonomous runtime adaptation mechanism. We make this choice deliberately since in domains such as healthcare and industrial control, restricting authorization can affect patient care and physical processes, so such changes should remain under accountable human control. After Z3 characterizes the feasible authorization configurations and their associated residual-risk/penalty trade-offs, administrators can select the configuration that best aligns with organizational priorities, operational constraints, and risk appetite. The latter can be expressed through organization-specific criteria, such as acceptable residual-risk thresholds, penalty budgets, or preferences over the risk--penalty trade-off. %

Because the recommended configurations depend only on the inputs of a context (its contextual factors, the method selected for it, and the domain model), they need not be computed at the moment of access. Administrators can compute and review them in advance for each recurring context, such as emergency access from an ambulance or remote access from an unmanaged device, and pre-approve one configuration per context. At runtime, the pre-approved configuration for the current context is enforced on top of the default authorization configuration $\pi_0$ by applying its operation restrictions and access-right adjustments. When contextual conditions change, or adaptive authentication selects a different method, the configuration approved for the new context applies, and the configurations are recomputed whenever the domain model or its parameters change. An organization willing to delegate the choice can instead fix an acceptable residual-risk threshold in advance, in which case the lowest-penalty recommended configuration that meets it can be applied automatically. Our prototype computes the recommended configurations but does not implement platform-specific policy propagation or enforcement; until a configuration is applied, the existing authorization policy remains in effect.

%% file: Evaluation.tex
\section{Evaluation}
We evaluate our approach to assess how effectively it mitigates residual risk while preserving access to the assets required for the most critical operations. To the best of our knowledge, prior research on adaptive authentication and adaptive authorization has not explicitly considered the \emph{residual risk} left by the selected authentication method, nor has it used that residual risk to drive fine-grained authorization adaptation. Our evaluation goal is therefore to determine whether using adaptive authorization as a second line of defense can meaningfully mitigate residual risk after authentication, without unnecessarily restricting permissions needed to execute the most critical operations.

The evaluation is based on formal modeling and scenario-based case studies, as no comprehensive public dataset exists for adaptive authentication \cite{bumiller_understanding_2023}. We use two case studies: (i) healthcare (Section~\ref{healthcare_case}) and (ii) industrial control systems (ICSs) (Section B of the supplementary material). We selected these case studies to demonstrate the applicability of our approach across domains with distinct contextual factors, assets, operational requirements, and security threats. The operations and associated assets listed in Table~\ref{tab:healthcare_operation_criticality_assets_all} and in Table~B.1 of the supplementary material define the default authorization configuration $\pi_0$ in the two case studies, in which every listed asset initially carries all four access rights.

Throughout the evaluation section, we refer to the adaptive authentication phase as Phase 1 and the adaptive authorization phase as Phase 2. The evaluation considers four dimensions. First, we assess the approach's \emph{effectiveness} in reducing residual risk while preserving access to the assets required for the most critical operations. Second, we compare it against standard \emph{authorization baselines} to examine whether it provides advantages over static or threshold-based authorization strategies. Third, we assess the \emph{robustness} of our approach under plausible uncertainty in attack-likelihood parameters, authentication methods' effectiveness, asset sensitivities, and risk aggregation choices. Fourth, we evaluate the \emph{computational overhead} of the approach to determine whether the analysis remains feasible in practice.

\paragraph{Baselines.} We compare our approach against three representative authorization baselines: \textit{Baseline A (Static RBAC)} grants the permissions assigned to the user’s role under a static RBAC policy after adaptive authentication selects an authentication method \cite{10.1145/501978.501980}. Authorization does not adapt to contextual changes or consider residual risk. \textit{Baseline B (Least-Privilege RBAC)} grants only the minimum RBAC permissions required for the task after adaptive authentication selects an authentication method~\cite{sandhu2000nist}. Although more restrictive, authorization remains static and is not optimized based on residual risk.
\textit{Baseline C (Context-threshold ABAC)} adapts authorization through predefined context-threshold rules (e.g., device, location) after adaptive authentication selects an authentication method~\cite{bailey_self-adaptive_2011,hu2014guide}. However, it does not account for the residual risk left by the selected authentication method.
\textit{Our approach}, in contrast, uses the residual risk left after adaptive authentication to enumerate admissible authorization configurations and evaluate them in terms of residual risk and functionality loss. We selected these baselines because they are widely used, representative authorization approaches, and provide meaningful points of comparison. The exact configuration instantiated for each baseline in each scenario is provided in the replication package~\footnote{\url{https://anonymous.4open.science/r/adaptive-auth-residual_risk-3BF9/}} as \path{parameters/baseline_configuration.csv}.

\paragraph{Implementation and setup.}

We implemented the experiments in Python using the Z3 Satisfiability Modulo Theories (SMT) solver version 4.13.3~\cite{10.1007/978-3-540-78800-3_24}, with the constraints encoded in SMT-LIB v2 syntax~\cite{Barrett}. Z3 has been widely evaluated for its scalability in both academic and industrial contexts, and it is often preferred for its rich API and flexible integration \cite{bjorner2019programming}. These properties make it a strong candidate for real-time symbolic reasoning tasks such as those found in adaptive security or authentication systems~\cite{backes2018semantic}. We encoded contextual factors, asset sensitivities, and operational criticality as inputs. We ran all experiments on the same machine: a 13-inch MacBook Pro (2022) with an Apple M2 chip (8-core CPU) and 8\, GB of RAM. All model parameters, including asset sensitivities and weights, attack likelihoods, mitigation coefficients, and operation weights, together with the SMT-LIB encodings and scripts, are available in the replication package.

\subsection{Healthcare Case Study} \label{healthcare_case}

We first evaluate the effectiveness of the approach in healthcare scenarios, focusing on whether adaptive authorization can reduce the residual risk left after authentication while preserving access to assets required for essential clinical operations.

\textbf{Outside the Hospital (S1)} represents remote access at an unusual time and over an insecure network, where a physician connects to the system from a caf\'e. %

\textbf{Ambulance Emergency Access (S2)} represents an emergency access situation in which a physician connects to the system from an ambulance during a critical case to retrieve the patient's profile, order lab tests, and adjust the care plan. %

\textbf{Late Night Access (S3)} represents an access situation in which a senior physician connects to the system from a personal laptop over an unsecured home Wi-Fi connection to review a deteriorating patient's lab results and diagnosis. %

In each scenario, the user can perform a set of operations with varying criticality and interact with assets with different sensitivity and access rights; Table~\ref{tab:healthcare_operation_criticality_assets_all} lists them.

\subsubsection{Results and Analysis}
Our approach returns a set of feasible authorization configurations rather than a single optimal one, since the preferred residual-risk/penalty trade-off can differ across organizations and risk appetites.
Each configuration corresponds to a point in the residual risk--penalty space, enabling security administrators to select a configuration aligned with their operational needs. Across healthcare scenarios, adaptive authentication selects different methods under different contexts, producing different residual-risk levels; adaptive authorization then enumerates alternative configurations at progressively lower residual-risk levels to reduce this residual risk. The step $\Delta$ controls how finely this trade-off space is explored: a smaller step yields more configurations and more solver calls, whereas a larger step yields fewer, more widely spaced ones. We use $\Delta = 0.012$ in all six scenarios, which yields between 11 and 21 recommended configurations per scenario, enough to capture each stage of the adaptation while keeping the set manageable for an administrator.

\begin{table*}[ht]
\caption{Phase 1 inputs and results for the healthcare scenarios (S1--S3).}
\centering
\footnotesize
\renewcommand{\arraystretch}{1.15}
\setlength{\tabcolsep}{3.5pt}
\begin{tabular}{|p{1cm}|p{3.9cm}|p{3cm}|p{1.25cm}|p{1.15cm}|p{2cm}|p{1cm}|}
\hline
\textbf{Scenario} &
\textbf{Contextual Factors} &
\textbf{Attack Likelihood} &
\textbf{Asset Value ($I(\pi_0)$)} &
\textbf{Total Risk} &
\textbf{Selected Method} &
\textbf{Residual Risk} \\
\hline

S1 &
Location=1; UnusualTime=1; InsecNetwork=1; UseMaxAgg=true &
PImpersAttack=L; PSessionAttack=H; PReplayAttack=M &
0.8251 & 0.5776 &
SmartCard + Fingerprint &
0.2063 \\
\hline

S2 &
Location=1; Emergency=1; UnsecuredWiFi=1; UseMaxAgg=true &
PImpersAttack=M; PSessionAttack=H; PReplayAttack=H & 0.8152 & 0.7337 &
SmartCard & 0.3261
 \\
\hline

S3 &
UnknownDevice=1; Location=1; UnsecuredWiFi=1; Scanner=0; Reader=0; UseMaxAgg=true &
PImpersAttack=M; PSessionAttack=M; PReplayAttack=M &
0.7691 & 0.4691 &
Token + Iris &
0.2692 \\
\hline
\end{tabular}

\vspace{4pt}
\footnotesize \textbf{Note:} L: $<0.55$; M: $[0.55, 0.65)$; H: $\geq 0.65$. InsecNetwork denotes an untrusted network in general; UnsecuredWiFi denotes an unsecured wireless network. UseMaxAgg indicates worst-case (max) aggregation across attacks.
\label{tab:healthcare_scenario_summary}
\end{table*}

Table~\ref{tab:healthcare_scenario_summary} shows the Phase 1 inputs and results for the healthcare scenarios. In \textbf{S1}, adaptive authentication selects \emph{SmartCard + Fingerprint} because remote access over an insecure network at an unusual time and location indicates elevated risk. After authentication, the residual risk under the default authorization configuration is $0.2063$, with an impact $I(\pi_0)$ of $0.8251$. We represent this starting point, $\pi_0$, as Config.~\#0, where all operations remain enabled, and all assets retain their original access rights. Since no adaptive authorization restriction has been applied yet, the authorization penalty is $0$.
Although SmartCard + Fingerprint is appropriate for this context, it leaves a residual risk, which adaptive authorization, as a second line of defense, aims to mitigate. In S1, our approach recommends 16 configurations. Table~\ref{tab:s1_configs} reports a representative subset; the full sets for all scenarios are in the replication package. In this and the other representative configuration tables, we include the default configuration (Config.~\#0), the configuration with the lowest residual risk, and configurations that illustrate each stage of adaptation: first tightening access rights alone, then disabling Optional and Nice-to-Have operations.
Config.~\#5 is the last configuration in which all operations remain enabled: it reduces the residual risk to $0.1564$ at a penalty of $0.0881$ entirely through fine-grained tightening of the access rights associated with selected assets. At lower residual-risk levels, our approach also disables operations, starting with the least critical, when doing so incurs a lower penalty than further tightening. Config.~\#6 is the first configuration in which an \textit{Optional} operation, \emph{AddEncounterNote}, is disabled. A second \textit{Optional} operation, \emph{AddReferralNote}, is disabled in Config.~\#8, and by Config.~\#12, all three \textit{Optional} operations are disabled, including \emph{GenerateReports}.
\textit{Nice-to-Have} operations remain enabled until the lowest residual-risk levels. \emph{PrescribeControlled} is disabled in Config.~\#15, while both \emph{PrescribeControlled} and \emph{PrescribeNonControlled} are disabled in Config.~\#16. In contrast, the two \textit{Must-Have} operations, \emph{DiagnoseMedicalConditions} and \emph{ViewLabResults}, remain enabled in every reported configuration. Overall, the residual risk decreases from $0.2063$ in the default authorization configuration to $0.0249$ in Config.~\#16, while the authorization penalty increases from $0$ to $0.8558$.

\begin{table*}[ht]
\caption{Representative authorization configurations in S1. Only enabled assets are listed, with their enabled access rights in parentheses.}
\centering
\footnotesize
\renewcommand{\arraystretch}{1.15}
\setlength{\tabcolsep}{3.5pt}
\begin{tabular}{|p{0.8cm}|p{1cm}|p{0.8cm}|p{3.9cm}|p{7.3cm}|}
\hline
\textbf{Config} & \textbf{Residual Risk} & \textbf{Penalty} & \textbf{Disabled Operations} & \textbf{Enabled Assets (Access Rights)} \\
\hline

\#0 &
0.2063 & 0.0000 &
None &
Diagnosis (R,W,U,D);
TreatmentPlan (R,W,U,D);
LabResults (R,W,U,D);
TestReports (R,W,U,D);
NonControlledPrescriptions (R,W,U,D);
PatientHistory (R,W,U,D);
ControlledPrescriptions (R,W,U,D);
PatientInteractions (R,W,U,D);
VisitSummaries (R,W,U,D);
ReferralNotes (R,W,U,D);
EncounterNotes (R,W,U,D);
PatientRecords (R,W,U,D). \\
\hline

\#5 &
0.1564 & 0.0881 &
None &
Diagnosis (R,W,U,D);
TreatmentPlan (R,W,U,D);
LabResults (R,W,U,D);
TestReports (R,W,U,D);
NonControlledPrescriptions (R,W,U,D);
PatientHistory (R,W,U,D);
ControlledPrescriptions (R);
PatientInteractions (R,W);
VisitSummaries (R,W);
ReferralNotes (R,W,U);
EncounterNotes (R,W,U,D);
PatientRecords (R). \\
\hline

\#6 &
0.1453 & 0.1298 &
AddEncounterNote &
Diagnosis (R,W,U,D);
TreatmentPlan (R,W,U,D);
LabResults (R,W,U,D);
TestReports (R,W,U,D);
NonControlledPrescriptions (R,W,U,D);
PatientHistory (R,W,U,D);
ControlledPrescriptions (R);
ReferralNotes (R,W,U,D);
EncounterNotes (R,W,U,D);
PatientRecords (R). \\
\hline

\#8 &
0.1215 & 0.2452 &
AddEncounterNote, AddReferralNote &
Diagnosis (R,W,U,D);
TreatmentPlan (R,W,U,D);
LabResults (R,W,U,D);
TestReports (R,W,U,D);
NonControlledPrescriptions (R,W,U);
PatientHistory (R,W,U);
ControlledPrescriptions (R);
PatientRecords (R). \\
\hline

\#12 &
0.0740 & 0.5192 &
AddEncounterNote, AddReferralNote, GenerateReports &
Diagnosis (R,W,U,D);
TreatmentPlan (R);
LabResults (R);
TestReports (R,W,U,D);
NonControlledPrescriptions (R,W,U,D);
PatientHistory (R);
ControlledPrescriptions (R). \\
\hline

\#15 &
0.0367 & 0.7596 &
PrescribeControlled, AddEncounterNote, AddReferralNote, GenerateReports &
Diagnosis (R);
TreatmentPlan (R);
LabResults (R);
TestReports (R,W,U,D);
NonControlledPrescriptions (R);
PatientHistory (R). \\
\hline

\#16 &
0.0249 & 0.8558 &
PrescribeControlled, PrescribeNonControlled, AddEncounterNote, AddReferralNote, GenerateReports &
Diagnosis (R);
TreatmentPlan (R);
LabResults (R);
TestReports (R,W). \\
\hline

\end{tabular}
\label{tab:s1_configs}
\end{table*}

In \textbf{S2}, adaptive authentication selects a \emph{SmartCard} under the emergency-access context, where the user accesses the system from an unusual location over unsecured Wi-Fi. After authentication, the residual risk under the default authorization configuration is $0.3261$, with an impact of $0.8152$. This starting point is Config.~\#0, in which all S2 operations (Table~\ref{tab:healthcare_operation_criticality_assets_all}) remain enabled and all assets retain their original access rights. Since no adaptive authorization restriction has been applied yet, the authorization penalty is $0$.
Adaptive authorization then identifies 21 recommended configurations with progressively lower residual-risk levels (Table~\ref{tab:s2_configs}). As in S1, the first stage of adaptation relies entirely on fine-grained restriction of access rights rather than disabling operations. Config.~\#5, for example, reduces the residual risk to $0.2766$ with a penalty of only $0.0660$ while retaining every operation, by restricting \emph{Diagnosis} and \emph{RadiologyImages} to Read-only access. Config.~\#14 is the last configuration in which all nine operations remain enabled, reducing the residual risk to $0.1683$ at a penalty of $0.3588$.
At lower residual-risk levels, our approach also disables operations, starting with the least critical. Config.~\#15 is the first configuration in which an \textit{Optional} operation, \emph{ViewRadiologyImages}, is disabled, reducing the residual risk to $0.1578$ at a penalty of $0.4051$. In Config.~\#16, \emph{ManageAppointment} is also disabled, further reducing the residual risk to $0.1460$, and by Config.~\#19 all three \textit{Optional} operations have been disabled, including \emph{PrintMedicalReport}. At that point, the approach also begins restricting the \textit{Nice-to-Have} operations, disabling \emph{RequestConsultation} and reducing the residual risk to $0.1095$ at a penalty of $0.6667$.
By Config.~\#21, the remaining \textit{Nice-to-Have} operations, \emph{CommunicateWithPatient} and \emph{ApproveDischargeSummary}, are also disabled, leaving only the three \textit{Must-Have} operations---\emph{AccessPatientProfile}, \emph{OrderLabTest}, and \emph{AdjustCarePlan}---enabled. Overall, the residual risk decreases from $0.3261$ to $0.0852$ while the penalty increases from $0$ to $0.8056$.

\begin{table*}[ht]
\caption{Representative authorization configurations in S2. Only enabled assets are listed, with their enabled access rights in parentheses.}
\centering
\footnotesize
\renewcommand{\arraystretch}{1.15}
\setlength{\tabcolsep}{3.5pt}
\begin{tabular}{|p{0.8cm}|p{1cm}|p{0.8cm}|p{3.7cm}|p{7.5cm}|}
\hline
\textbf{Config} &
\textbf{Residual Risk} &
\textbf{Penalty} &
\textbf{Disabled Operations} &
\textbf{Enabled Assets (Access Rights)} \\
\hline

\#0 &
0.3261 & 0.0000 &
None &
Diagnosis (R,W,U,D);
TreatmentPlan (R,W,U,D);
LabResults (R,W,U,D);
TestReports (R,W,U,D);
PatientHistory (R,W,U,D);
PatientInteractions (R,W,U,D);
VisitSummaries (R,W,U,D);
ReferralNotes (R,W,U,D);
EncounterNotes (R,W,U,D);
PatientRecords (R,W,U,D);
PatientDemographics (R,W,U,D);
RadiologyImages (R,W,U,D);
DischargeSummaries (R,W,U,D);
PatientCommunications (R,W,U,D). \\
\hline

\#5 & 0.2766 & 0.0660 & None & Diagnosis (R); TreatmentPlan (R,W,U,D); LabResults (R,W,U,D); TestReports (R,W,U,D); PatientHistory (R,W,U,D); PatientInteractions (R,W,U,D); VisitSummaries (R,W,U,D); ReferralNotes (R,W,U,D); EncounterNotes (R,W,U,D); PatientRecords (R,W,U,D); PatientDemographics (R,W,U,D); RadiologyImages (R); DischargeSummaries (R,W,U,D); PatientCommunications (R,W,U,D).

 \\
\hline

\#14 & 0.1683 & 0.3588 & None & Diagnosis (R); TreatmentPlan (R,W,U); LabResults (R,W); TestReports (R,W); PatientHistory (R); PatientInteractions (R); VisitSummaries (R,W); ReferralNotes (R,W); EncounterNotes (R,W); PatientRecords (R,W,U,D); PatientDemographics (R); RadiologyImages (R); DischargeSummaries (R,W,U); PatientCommunications (R,W,U,D). \\ \hline

\#15 &
0.1578 & 0.4051 &
ViewRadiologyImages &
Diagnosis (R);
TreatmentPlan (R,W,U);
LabResults (R,W);
TestReports (R);
PatientHistory (R);
PatientInteractions (R);
VisitSummaries (R,W);
ReferralNotes (R,W);
EncounterNotes (R,W);
PatientRecords (R,W,U,D);
PatientDemographics (R);
DischargeSummaries (R,W,U);
PatientCommunications (R,W,U,D). \\
\hline

\#16 &
0.1460 & 0.4745 &
ViewRadiologyImages, ManageAppointment &
Diagnosis (R);
TreatmentPlan (R,W,U);
LabResults (R,W);
TestReports (R);
PatientHistory (R);
PatientInteractions (R);
ReferralNotes (R,W);
EncounterNotes (R,W);
PatientRecords (R,W,U,D);
PatientDemographics (R);
DischargeSummaries (R,W,U);
PatientCommunications (R,W). \\
\hline

\#19 &
0.1095 & 0.6667 &
RequestConsultation, ViewRadiologyImages,
ManageAppointment, PrintMedicalReport &
Diagnosis (R);
TreatmentPlan (R,W,U);
LabResults (R,W);
TestReports (R);
PatientHistory (R);
PatientInteractions (R,W);
PatientRecords (R);
PatientDemographics (R);
DischargeSummaries (R,W,U);
PatientCommunications (R,W,U,D). \\
\hline

\#21 &
0.0852 & 0.8056 &
ApproveDischargeSummary, RequestConsultation,
CommunicateWithPatient, ViewRadiologyImages,
ManageAppointment, PrintMedicalReport &
Diagnosis (R);
TreatmentPlan (R,W,U);
LabResults (R,W);
TestReports (R);
PatientHistory (R);
PatientRecords (R);
PatientDemographics (R,W,U). \\
\hline

\end{tabular}
\label{tab:s2_configs}
\end{table*}

In \textbf{S3}, adaptive authentication selects \emph{Token + Iris} under a context characterized by an unfamiliar location, unsecured Wi-Fi, and an unknown device. After authentication, the residual risk under the default authorization configuration is $0.2692$, with an impact of $0.7691$. This starting point is represented by Config.~\#0, in which all S3 operations (Table~\ref{tab:healthcare_operation_criticality_assets_all}) remain enabled, and all assets retain their original access rights.
Adaptive authorization then identifies 20 recommended configurations associated with progressively lower residual-risk levels, a representative subset of which is reported in Table~A.5 of the supplementary material. As in S1 and S2, the reduction in residual risk is initially achieved entirely by tightening access rights for selected assets. Config.~\#6 is the last configuration in which all seven operations remain enabled: it reduces the residual risk from $0.2692$ to $0.2076$ at a penalty of $0.0881$, by reducing \emph{ControlledPrescriptions} to \textit{Read-only} access and tightening the access rights associated with \emph{PatientInteractions}, \emph{VisitSummaries}, \emph{ReferralNotes}, and \emph{PatientRecords}.
At lower residual-risk levels, our approach also disables operations, starting with the least critical. Config.~\#7 is the first configuration in which we disable an \textit{Optional} operation, \emph{AddEncounterNote}, reducing the residual risk to $0.1966$ at a penalty of $0.1266$. In Config.~\#9, the second \textit{Optional} operation, \emph{AddReferralNote}, is also disabled, reducing the residual risk to $0.1691$. The remaining \textit{Optional} operation, \emph{GenerateReports}, remains available for several further configurations while the access rights on its assets are progressively restricted. By Config.~\#16, all three \textit{Optional} operations are disabled. At this point, the residual risk has decreased to $0.0875$ and the authorization penalty has increased to $0.5577$, while the two \textit{Must-Have} and two \textit{Nice-to-Have} operations remain enabled.
Further reductions require progressively restricting the \textit{Nice-to-Have} operations. In Config.~\#19, \emph{PrescribeControlled} is disabled, leaving \emph{PrescribeNonControlled} available alongside the two \textit{Must-Have} operations and reducing the residual risk to $0.0524$ at a penalty of $0.7468$. Finally, Config.~\#20 also disables \emph{PrescribeNonControlled}, leaving only the \textit{Must-Have} operations \emph{DiagnoseMedicalConditions} and \emph{ViewLabResults} enabled, with a residual risk of $0.0398$ and a penalty of $0.7981$.

These results show that further reduction in residual risk comes at the cost of increasingly restrictive authorization. The order of adaptation is the same in all three scenarios: access rights are tightened first, \textit{Nice-to-Have} operations are disabled only once all \textit{Optional} operations have been disabled, and the \textit{Must-Have} operations remain enabled in every recommended configuration. This order largely follows from the model, since Must-Have operations are protected by constraint (C1) and Optional operations carry the smallest weights, which makes them the cheapest to disable.
In each scenario, the residual risk is reduced substantially before any operation is disabled, by about $24\%$ in S1 (from $0.2063$ to $0.1564$), $48\%$ in S2 (from $0.3261$ to $0.1683$), and $23\%$ in S3 (from $0.2692$ to $0.2076$) relative to the initial residual risk, suggesting that administrators can obtain meaningful risk reduction at little functional cost before facing the steeper trade-offs of operation disablement.

\begin{figure*}[ht]
 	\centering
 	\begin{subfigure}[b]{0.33\linewidth}
 		\includegraphics[width=\linewidth]{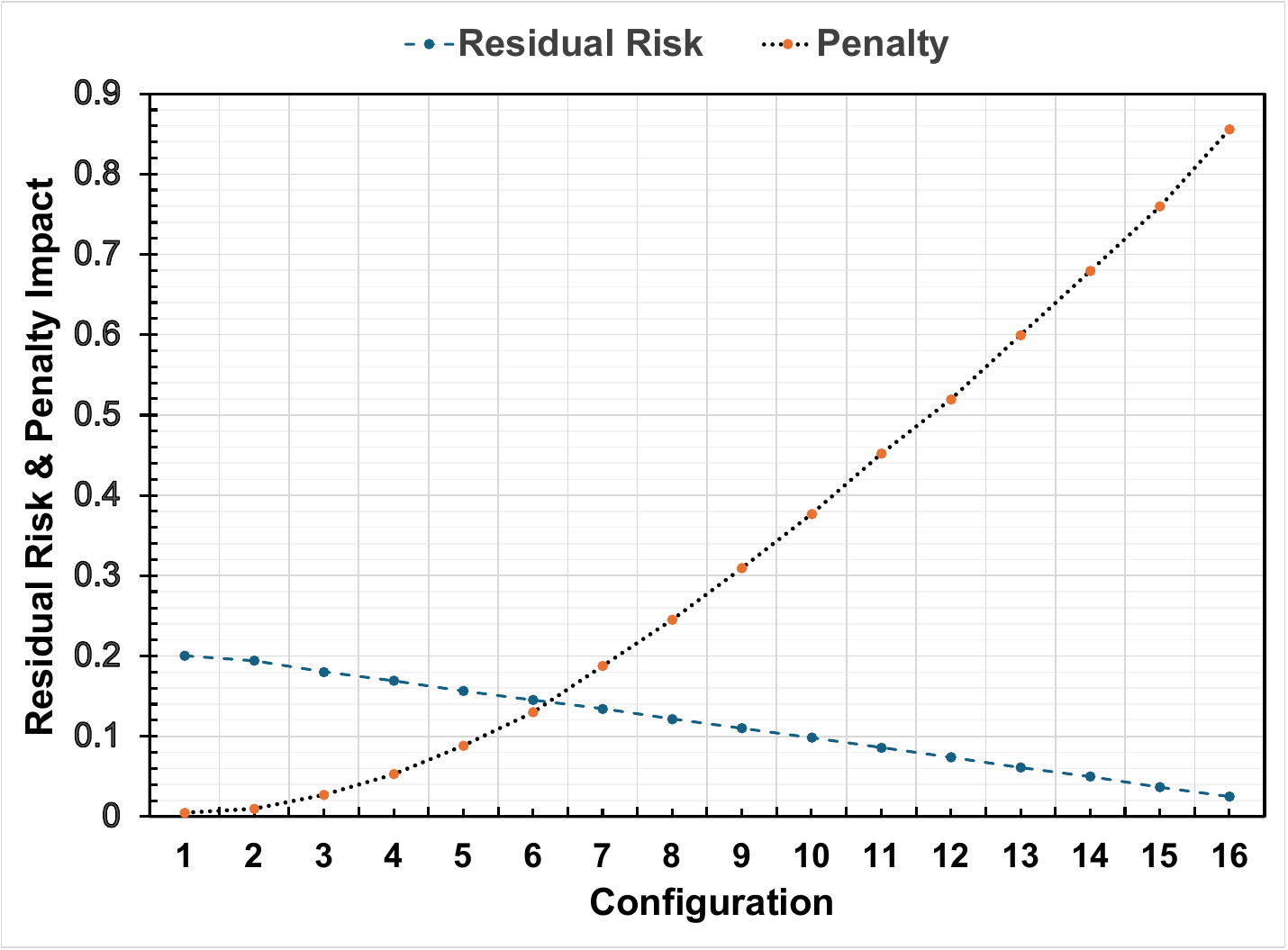}
 		\caption{S1 (Outside the Hospital)}
 		\label{fig-RRvP-a}
 	\end{subfigure}
 	\begin{subfigure}[b]{0.33\linewidth}
 		\includegraphics[width=\linewidth]{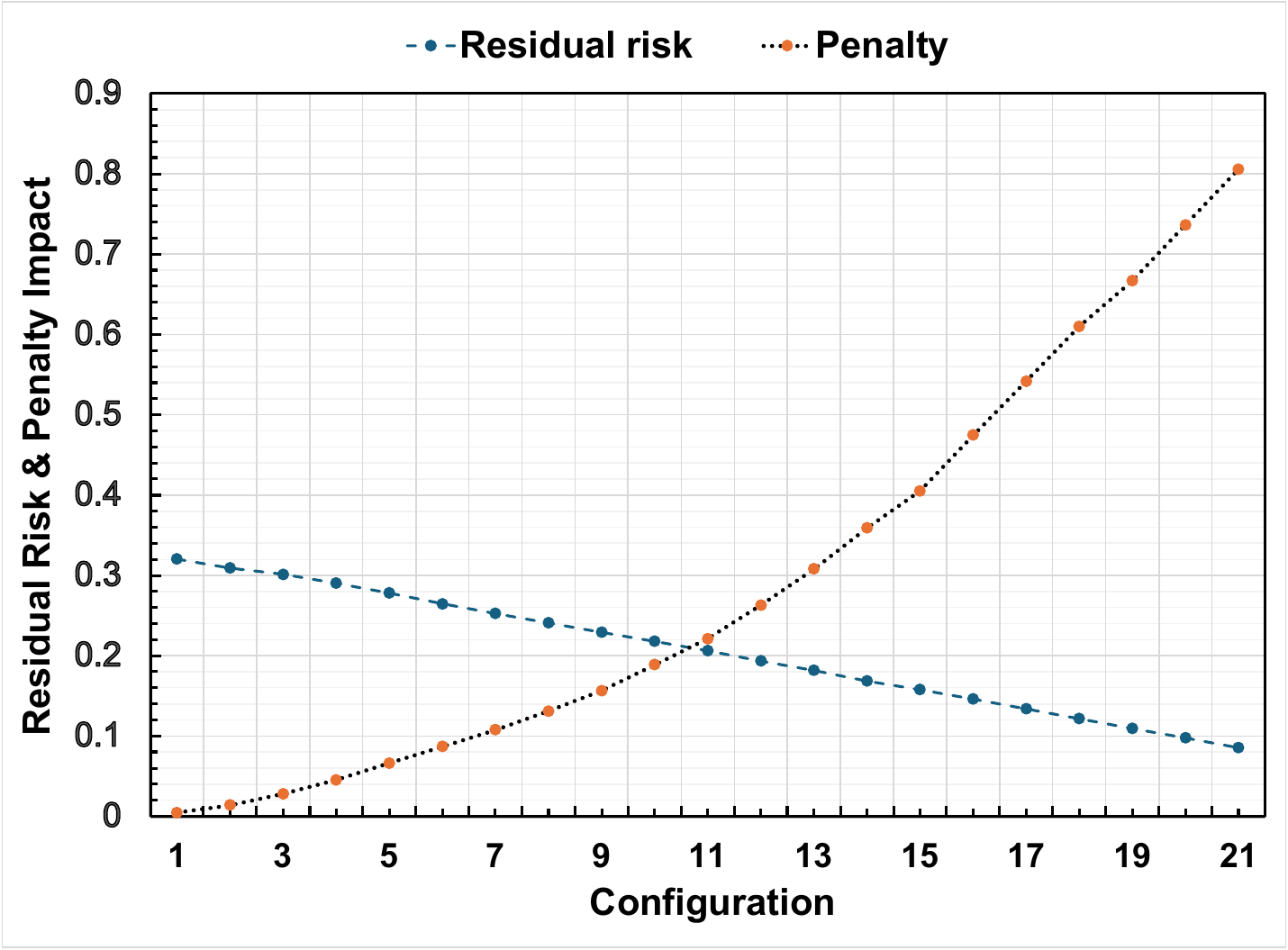}
 		\caption{S2 (Ambulance Emergency Access)}
 		\label{fig-RRvP-b}
 	\end{subfigure}
 \begin{subfigure}[b]{0.33\linewidth}
 	\includegraphics[width=\linewidth]{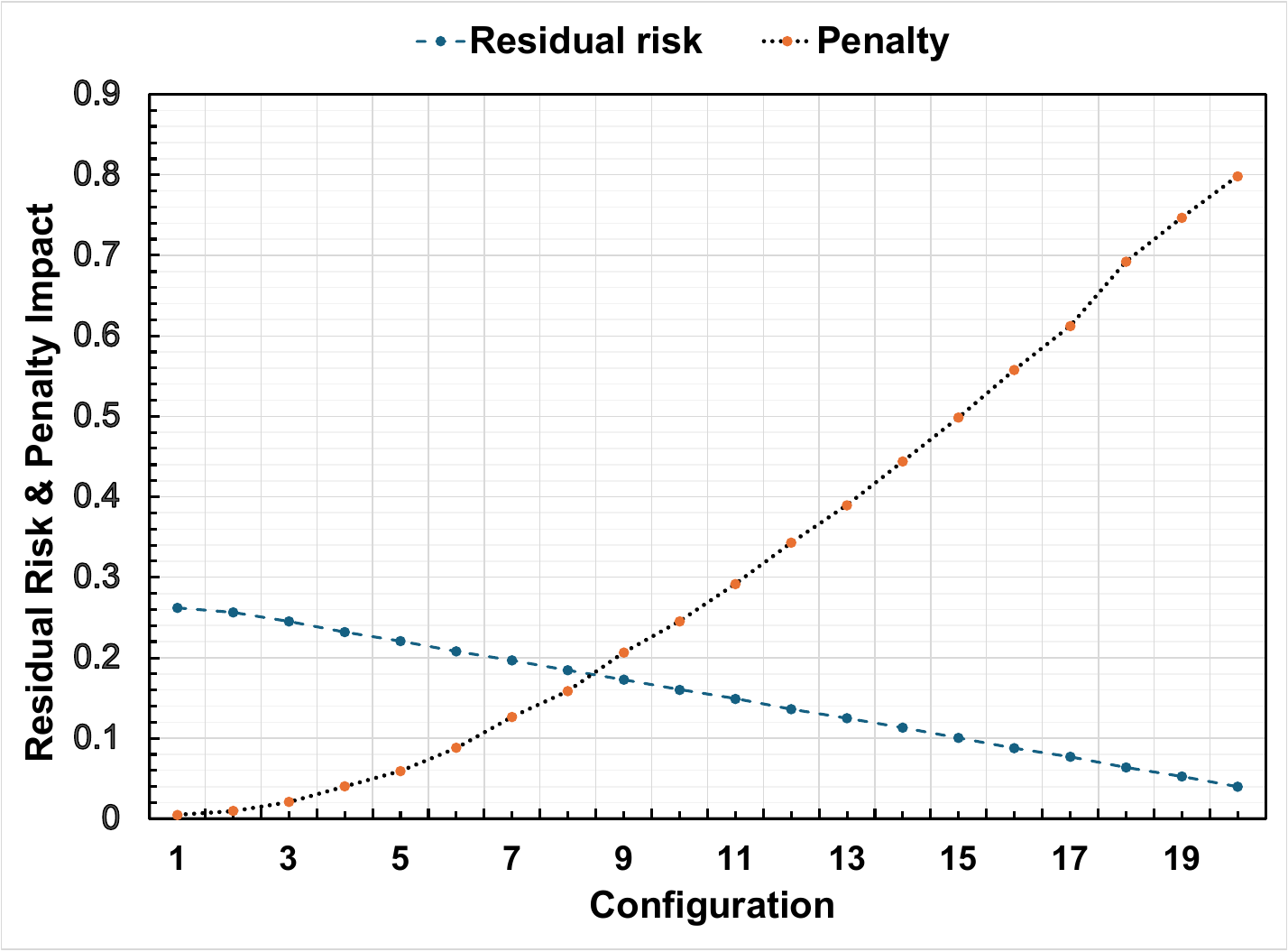}
 	\caption{S3 (Late Night Access)}
 	\label{fig-RRvP-c}
 \end{subfigure}

 	\caption{Residual risk vs. authorization penalty across scenarios S1--S3. Each point represents a feasible authorization configuration.} \Description{Three line charts, one each for scenarios S1, S2, and S3, plotting residual risk and authorization penalty (both on a 0 to 0.9 scale) against the configuration number. In every scenario, residual risk decreases steadily across configurations while the penalty rises from zero, slowly at first and then more steeply, so the two curves cross partway along and the last configurations buy small risk reductions at a large penalty.} \Description{Three line charts, one each for scenarios S1, S2, and S3, plotting residual risk and authorization penalty (both on a 0 to 0.9 scale) against the configuration number. In every scenario, residual risk decreases steadily across configurations while the penalty rises from zero, slowly at first and then more steeply, so the two curves cross partway along and the last configurations buy small risk reductions at a large penalty.}
 	\label{fig:RRvP}
 \end{figure*}

\paragraph{Residual Risk vs Penalty.}
Figure~\ref{fig:RRvP}(a--c) shows the relationship between residual risk (RR) and penalty across feasible authorization configurations produced by our approach for the healthcare scenarios (16 in S1, 21 in S2, and 20 in S3). In all three scenarios, residual risk decreases as the authorization policy becomes more restrictive, while the penalty generally increases. This indicates that stronger residual risk mitigation is achievable, but at the cost of increasingly restrictive authorization. In particular, targeted tightening of access rights on high-impact assets can reduce risk more efficiently than disabling entire operations: in S2, the residual risk is nearly halved before any operation is disabled. To illustrate what these trade-offs mean in practice, consider two configurations from S1. Config.~\#6 reduces the residual risk by $30\%$ relative to $\pi_0$, from $0.2063$ to $0.1453$, while disabling only one \textit{Optional} operation, at a penalty of $0.1298$. Config.~\#16 reduces it by $88\%$, to $0.0249$, but disables five of the seven operations, leaving only the two \textit{Must-Have} operations, at a penalty of $0.8558$. An administrator whose acceptable residual risk is $0.15$ would therefore select Config.~\#6, the first configuration to meet that threshold, and retain almost all functionality, whereas reaching the lowest residual-risk level requires giving up all \textit{Optional} and \textit{Nice-to-Have} operations. S2 and S3 exhibit the same pattern.

Across the three scenarios, the largest marginal reductions in residual risk tend to occur in earlier configurations, whereas further improvements near the lowest-risk region require disproportionately larger penalty increases, indicating diminishing returns near the extreme low-risk end. More importantly, the plots expose a range of risk--penalty trade-offs and motivate reporting multiple candidate configurations so security administrators can select the lowest-penalty configuration that satisfies their desired residual-risk thresholds, rather than minimizing residual risks at any cost.

\paragraph{Per-attack residual risks.}

We also break down residual risk by attack type (impersonation, replay, and session hijacking) to show which attack determines the overall residual risk in each scenario, and therefore where authorization adaptation stops being effective. Because all three attacks share the same impact (Section~\ref{residual-risk}), each per-attack residual risk is its post-authentication likelihood scaled by $I(\pi)$. Tightening authorization therefore reduces all three in the same proportion, and their relative ordering does not change across configurations. The scenarios differ in which attack dominates: in S1, replay risk remains the highest, followed by impersonation and then session hijacking; in S2, replay and session-hijacking risks are equal and remain higher than impersonation risk; and in S3, replay risk remains the highest, followed by impersonation, while session-hijacking risk is substantially lower. Figure~A.1 of the supplementary material plots the residual risk of each attack for every recommended configuration in S1--S3.

Because the healthcare scenarios use max-based aggregation, the overall residual risk follows the highest per-attack curve at each configuration, so the dominant attack determines it throughout. In S1 and S3, replay dominates, whereas in S2, replay and session hijacking are tied. The dominant attack therefore sets a lower bound on what authorization adaptation can achieve in each scenario: an administrator whose acceptable residual risk falls below that bound cannot reach it by restricting authorization further, but only by strengthening authentication against that attack. This complements the risk--penalty trade-offs in Figure~\ref{fig:RRvP}, which show what each residual-risk level costs in functionality.

\subsubsection{Baselines and Results}
\label{sec:healthcare_baselines}

We compare our approach against the three baselines introduced at the beginning of Section 4. \textbf{Static RBAC} grants the full set of role-defined operations and access rights \cite{10.1145/501978.501980}; \textbf{Least-Privilege RBAC} enables only the minimum operations and assets required for the role \cite{sandhu2000nist}; and \textbf{Context-threshold ABAC} maps contextual attributes through predefined thresholds to LOW, MEDIUM, or HIGH authorization tiers, providing a representative context-aware comparison rather than the full expressiveness of ABAC \cite{bailey_self-adaptive_2011,hu2014guide}. All baselines are evaluated using the \emph{same} residual risk model, attack set (impersonation, replay, session hijacking), and risk aggregation. The only difference is the authorization configuration~$\pi$ produced by each baseline, ensuring that differences in residual risk and penalty reflect authorization decisions rather than changes in the underlying risk model.

Table~\ref{tab:baseline_healthcare} summarizes the baseline results for the three healthcare scenarios. \textit{Static RBAC} yields the lowest penalty but the highest residual risk, since it preserves broad access rights and therefore leaves a higher authorization-dependent impact $I(\pi)$. By contrast, \textit{least-privilege RBAC} consistently reduces residual risk, but at a higher penalty due to reductions in accessible functionality. \textit{ABAC} reduces residual risk further, but this stronger restriction also results in the highest penalty among the baselines. In \textbf{S1} and \textbf{S3}, ABAC selects the HIGH tier, resulting in the strongest restriction and the lowest baseline residual risk. In \textbf{S2}, ABAC selects the MEDIUM tier and preserves more functionality than a HIGH-tier policy, while still reducing residual risk more than either static or least-privilege RBAC.

Because our approach returns multiple feasible authorization configurations whereas each baseline yields a single configuration,  Table~\ref{tab:baseline_healthcare} reports our approach as a range of residual risks and penalties. This range cannot be compared directly against a baseline's single point, since it overlaps every baseline and whether our approach appears better depends on which end of it is read. We therefore also compare them in terms of residual-risk/penalty trade-offs. More specifically, for each baseline point, Table~\ref{tab:baseline_tradeoff_healthcare} reports: (i) the best residual risk achieved by our approach at the same or lower penalty, (ii) the lowest penalty at which our approach reaches the same or lower residual risk, and (iii) whether any non-default configuration produced by our approach Pareto-dominates the baseline, that is, achieves no higher residual risk and no higher penalty, with at least one of the two strictly lower. We use dominance because the two metrics trade off against each other: a lower residual risk obtained at a higher penalty does not by itself demonstrate an improvement over the baseline.

\begin{table*}[ht]
\centering
\caption{Baseline results for healthcare scenarios S1--S3.}
\label{tab:baseline_healthcare}
\footnotesize
\begin{tabular}{llcccc}
\toprule
Scenario &
Baseline &
Authorization Tier &
Initial Residual Risk &
Residual Risk After Authorization &
Penalty (functionality loss) \\
\midrule

S1 & Static RBAC
& --
& 0.206
& 0.206
& 0.000 \\

S1 & Least-privilege RBAC
& --
& 0.206
& 0.138
& 0.231 \\

S1 & ABAC
& HIGH
& 0.206
& 0.055
& 0.712 \\

S1 & \textbf{Our Approach}
& --
& 0.206
& \textbf{0.206--0.025}
& \textbf{0.000--0.856} \\

\midrule

S2 & Static RBAC
& --
& 0.326
& 0.326
& 0.000 \\

S2 & Least-privilege RBAC
& --
& 0.326
& 0.288
& 0.167 \\

S2 & ABAC
& MEDIUM
& 0.326
& 0.283
& 0.181 \\

S2 & \textbf{Our Approach}
& --
& 0.326
& \textbf{0.326--0.085}
& \textbf{0.000--0.806} \\

\midrule

S3 & Static RBAC
& --
& 0.269
& 0.269
& 0.000 \\

S3 & Least-privilege RBAC
& --
& 0.269
& 0.182
& 0.231 \\

S3 & ABAC
& HIGH
& 0.269
& 0.029
& 0.885 \\

S3 & \textbf{Our Approach}
& --
& 0.269
& \textbf{0.269--0.040}
& \textbf{0.000--0.798} \\

\bottomrule
\end{tabular}
\end{table*}

The trade-off-based comparison shows that our approach provides favorable trade-offs relative to the baselines in several healthcare scenarios. In S1, our approach Pareto-dominates both least-privilege RBAC and ABAC, achieving lower residual risk at a lower penalty.
In S2, our approach also Pareto-dominates both least-privilege RBAC and ABAC. Consider least-privilege RBAC, which leaves a residual risk of $0.288$ at a penalty of $0.167$. For the same functionality loss, our approach reaches a residual risk of $0.229$, about $20\%$ lower; alternatively, it reaches the baseline's residual risk at a penalty of only $0.066$, about $60\%$ less functionality loss. Our approach achieves this residual-risk reduction at a lower penalty than least-privilege RBAC because it first tightens access rights without disabling any operation, whereas least-privilege RBAC disables multiple operations to reach the same residual-risk level. An administrator can therefore either obtain more risk reduction for the same cost, or the same risk reduction at a lower cost, because the configurations are derived from the residual risk left by the selected authentication method rather than from a fixed policy.
In S3, our approach Pareto-dominates least-privilege RBAC, achieving a lower residual risk ($0.169$ versus $0.182$) at a lower penalty ($0.207$ versus $0.231$). However, it does not reach the very low residual-risk level achieved by ABAC ($0.029$); the lowest residual risk our approach achieves is $0.040$, although at a lower penalty than the ABAC configuration. Static RBAC is not Pareto-dominated in any of the scenarios because it has zero penalty. The default Config.~\#0 produced by our approach corresponds to this unrestricted authorization configuration, but it is excluded from the trade-off dominance analysis because the comparison focuses on non-default authorization adaptations.

\begin{table*}[ht]
\centering
\caption{Trade-off-based comparison with baselines for healthcare scenarios. %
}
\label{tab:baseline_tradeoff_healthcare}
\footnotesize
\begin{tabular}{llccccc}
\toprule
Scenario &
Baseline &
Baseline RR &
Baseline P &
Best Our RR at $P \leq P_b$ &
Lowest Our P at $RR \leq RR_b$ &
Pareto-dominant? \\
\midrule

S1 & Static RBAC
& 0.206
& 0.000
& --
& --
& No \\

S1 & Least-privilege RBAC
& 0.138
& 0.231
& 0.134
& 0.188
& Yes \\

S1 & ABAC
& 0.055
& 0.712
& 0.050
& 0.679
& Yes \\

\midrule

S2 & Static RBAC
& 0.326
& 0.000
& --
& --
& No \\

S2 & Least-privilege RBAC
& 0.288
& 0.167
& 0.229
& 0.066
& Yes \\

S2 & ABAC
& 0.283
& 0.181
& 0.229
& 0.066
& Yes \\

\midrule

S3 & Static RBAC
& 0.269
& 0.000
& --
& --
& No \\

S3 & Least-privilege RBAC
& 0.182
& 0.231
& 0.169
& 0.207
& Yes \\

S3 & ABAC
& 0.029
& 0.885
& 0.040
& --
& No \\

\bottomrule
\end{tabular}
\end{table*}

A key observation is that the baselines do not use the residual risk left by the selected authentication method to adapt authorization. Static RBAC is fixed by design, least-privilege RBAC applies a fixed reduced-privilege policy, and ABAC adapts through predefined context tiers while remaining decoupled from method-specific residual risk. In contrast, our approach explicitly uses the residual risk left after authentication to drive fine-grained authorization adaptation. After adaptive authentication selects a method, we quantify the residual risk under that method and context, and then enumerate feasible authorization configurations that reduce the authorization-dependent impact $I(\pi)$ and, consequently, the residual risk.
Rather than returning a single rigid decision, the approach exposes multiple admissible configurations, allowing administrators to select the one that best aligns with their risk appetite and operational constraints. Overall, the baseline comparison shows that explicitly using post-authentication residual risk enables finer-grained and more flexible authorization adaptation than static or threshold-based strategies.

\subsubsection{Sensitivity Analysis}
\label{sec:healthcare_sensitivity}

Because attack likelihoods, mitigation coefficients, and asset sensitivities are derived from the literature and expert judgment rather than empirical measurement, we assessed robustness through a one-factor-at-a-time sensitivity analysis on the healthcare scenarios, perturbing each of these parameters individually and also varying the risk aggregation function and the penalty weight $\beta$ ($0.70$--$0.99$).
Residual-risk values changed with the perturbations, most strongly when a perturbation changed the dominant attack, and least under asset-sensitivity perturbations, but the decisions remained stable: the selected authentication method was unchanged apart from tied alternatives in S3, and under the parameter perturbations the lowest-penalty recommendation disabled no operation. Because this recommendation is the first restriction step, the analysis does not assess whether configurations at lower residual-risk levels change. Replacing Max with Average aggregation lowered the residual risk without changing the selected authentication method, and the lowest-penalty recommendation tightened access rights further without disabling any operation; aggregation therefore had the largest systematic effect and should be chosen according to the risk interpretation of the domain. Varying $\beta$ changed the penalty values and some intermediate configurations, but not the selected authentication method or the achievable residual-risk range. Due to page limits, the full sensitivity analysis is reported in Section~A.1 of the supplementary material.

\subsubsection{Overhead}
To assess the approach's practical feasibility, we measure its computational overhead in healthcare scenarios. We run the pipeline 100 times for each scenario, %
and because each phase invokes Z3 multiple times%
, we report the overhead \emph{separately} for the two phases to isolate the cost of authentication selection from the cost of authorization adaptation.
For each phase, we report the mean across 100 runs for end-to-end phase runtime and memory usage. %
Table~\ref{tab:Healthcare-Overhead} summarizes the overhead of Phase 1 and Phase 2 for scenarios S1–S3. The results show that the computational overhead varies across scenarios, particularly in Phase~2, where Z3 explores multiple residual-risk intervals to generate alternative authorization configurations. Memory usage remains relatively stable across the healthcare scenarios, while runtime differences reflect scenario-specific variations in the solver search effort. Overall, these results show that the computational cost depends not only on the size of the SMT encoding but also on the constraints and feasible authorization configurations associated with each scenario.

To further characterize the complexity of the SMT problems solved for the healthcare scenarios, we report both encoding-size and solver-effort metrics (Table~\ref{tab:Healthcare-Overhead}). For encoding size, we report: (i) the number of declared variables (i.e., \texttt{declare-fun} and \texttt{declare-const} statements), (ii) the number of asserted constraints (\texttt{assert} statements), (iii) the number of arithmetic/comparison operator occurrences (\texttt{arith}; e.g., \texttt{+}, \texttt{-}, \texttt{*}, \texttt{/}, \texttt{<=}, \texttt{>=}), and (iv) the number of conditional expressions (\texttt{ite}; occurrences of \texttt{(ite ...)}). For solver effort, we report standard Z3 statistics, including the number of branching decisions, the number of propagated literals, and the number of solver restarts. Solver-effort statistics are aggregated over all solver calls executed within each phase and then averaged over the 100 runs (hence non-integer means can occur). Together, these metrics provide insight into both the structural size of the SMT encoding and the effort Z3 requires to compute feasible configurations, complementing the measured runtime and memory results.

\begin{table*}[htbp]
\centering
\caption{Overhead and encoding characteristics for the healthcare scenarios (Phase 1 and Phase 2 reported separately).}
\label{tab:Healthcare-Overhead}
\footnotesize
\begin{tabular}{llrrrrrrrrrr}
\toprule
Scenario & Phase & Runs & Time (ms) & Memory (MB) &
decisions & propagations & restarts & variables & asserts & arith & ite \\
\midrule
S1 & P1 & 100 & 5822.09  & 19.54 & 4253.65  & 17423.12  & 2.60  & 174 & 243 & 771 & 120 \\
S1 & P2 & 100 & 21821.65 & 39.66 & 5206.76  & 176581.74 & 11.50 & 155 & 211 & 435 & 45  \\
\midrule
S2 & P1 & 100 & 6974.74  & 19.47 & 15546.20 & 41879.27  & 4.62  & 189 & 270 & 842 & 134 \\
S2 & P2 & 100 & 69372.98 & 41.60 & 7932.88  & 390171.91 & 28.67 & 173 & 223 & 446 & 49  \\
\midrule
S3 & P1 & 100 & 5759.53  & 19.50 & 4664.00  & 17426.44  & 2.50  & 175 & 248 & 778 & 124 \\
S3 & P2 & 100 & 26203.62 & 40.19 & 4684.06  & 199559.55 & 12.29 & 155 & 203 & 400 & 48  \\
\bottomrule
\end{tabular}
\end{table*}

The overhead results distinguish between the cost of individual solver queries and that of generating the complete set of recommendations. For example, in S1, Phase~1 requires 18 solver calls and takes an average of 5.82\,s in total, or about 323\,ms per solver call. Phase~2 requires 54 solver calls and takes 21.82\,s to explore the residual-risk intervals and generate the complete set of authorization configurations, corresponding to approximately 404\,ms per solver call. Thus, the majority of the end-to-end cost results from repeatedly solving the SMT encoding to construct the full risk--penalty trade-off space, rather than from a single authorization decision. The current prototype therefore targets exhaustive generation and analysis of feasible authorization configurations rather than latency-optimized online deployment. In an operational setting, the complete trade-off space could be precomputed or cached where the relevant model parameters remain unchanged, or Z3 could be queried only for the residual-risk level required by the current decision. Optimizing these mechanisms is left for future work.

To assess generalizability beyond healthcare, we evaluated three ICS scenarios: Remote Maintenance (S4), involving access from a new location at an unusual time using an unfamiliar device; On-Site Emergency Operation (S5), requiring emergency access via a shared tablet; and High-Privilege Administrative Access (S6), involving remote access over an untrusted network. Unlike healthcare, the ICS scenarios use Average rather than Max aggregation to evaluate the approach under a different risk aggregation function, not because Average is preferable for ICS; Max remains applicable without modifying the model. The approach mitigates residual risk while preserving critical operations by tightening access rights first, disabling Nice-to-Have operations only after all Optional operations, and retaining all Must-Have operations. In S5, adaptive authentication selects \emph{SmartCard}, leaving a residual risk of $0.1547$, and our approach recommends 11 configurations, from which administrators choose according to their risk appetite. Our approach Pareto-dominates least-privilege RBAC and ABAC across all three scenarios, with lower Phase~2 runtimes (3.3--10.8\,s) than in healthcare. Detailed scenarios, operation classifications, results, baseline comparisons, and overhead are provided in Section~B of the supplementary material.

\subsection{Threats to Validity}
\label{threat_to_validaty}
\paragraph{Internal validity.} The correctness of our evaluation depends on how accurately contextual factors, authentication vulnerabilities, and asset sensitivities are represented in the model. We derived some of these values from expert knowledge and the literature rather than empirical measurement, which may introduce bias. We mitigated this threat by validating all Z3 encodings, cross-checking the logical consistency between authentication, authorization, and residual risk constraints, and performing multiple experimental sweeps under different configurations to ensure stable results. The sensitivity analysis (Section~\ref{sec:healthcare_sensitivity}) perturbs one parameter at a time and tracks the least-restrictive recommendation, so it does not capture interactions between parameters or changes in the more restrictive recommendations.

\paragraph{Construct validity.}
Our constructs for residual risk, penalty, and authorization are formal abstractions that simplify real system behaviors. While these abstractions enable formal reasoning, they may overlook runtime aspects such as user response delays, session expirations, or dynamic threat evolution. To address this, we grounded our approach in established risk definitions and verified that computed values behaved consistently across different scenarios. However, the authorization configurations generated by the approach have not been independently assessed by healthcare or ICS domain experts, nor have they been validated through realistic attack simulations. Future empirical validation involving domain experts, real authentication logs, and attack simulations would further strengthen confidence in the model and its generated authorization decisions. The authorization penalty treats restriction as a loss of functionality. In safety-critical settings such as S2 and S5, however, disabling an operation or tightening access rights may itself create safety risk, for instance by preventing a clinician from adjusting a care plan during an emergency. Our model partly accounts for this by never disabling Must-Have operations, but it does not quantify the safety consequences of restricting Nice-to-Have operations. Modeling such consequences explicitly, or pairing restrictions with emergency override mechanisms, is an important direction for future work.

\paragraph{External validity.} Our evaluation focused on two domains, healthcare and ICSs, each involving highly sensitive assets and strong contextual variability.
Results may not generalize to other domains such as finance or the Internet of Vehicles, where assets and contextual risks differ. To enhance generalizability, we plan to apply the model to additional domains and investigate transfer learning and reinforcement-learning-based tuning to automatically adjust contextual and residual risk parameters. While Z3 provides deterministic reasoning, its scalability on large configurations remains limited. Future work will explore parallel and hybrid optimization to improve runtime efficiency. Additionally, we selected the attacks considered in this evaluation (impersonation, replay, and session hijacking) because they represent common authentication and session-related threats in networked, data-sensitive environments~\cite{alaca_device_2016,wang_attacks_2021}. However, this set is not exhaustive, as other attack types (e.g., insider misuse, privilege escalation, denial-of-service, or timing attacks) could also affect authorization decisions. This limitation may influence the completeness of the evaluation, although the selected attacks cover the most frequent and impactful categories relevant to residual risks after authentication. Future work will expand the attack set to include a wider range of threat scenarios to strengthen generalizability.

%% file: Related_Work.tex
\section{Related Work}
\label{sec:related_work}

\paragraph{Authentication risks and adaptive authentication}
Authentication systems face attacks against credentials, authenticators, and the environments in which they operate~\cite{fi15040146,wang_attacks_2021}.
Wang et al.~\cite{wang_attacks_2021} survey attacks against password- and biometric-based systems and compare the strengths and limitations of corresponding defenses.
Chaudhary et al.~\cite{10.1016/j.cosrev.2019.03.002} explain that password managers reduce the burden of remembering credentials, while concentrating sensitive information in an attractive target for attackers; they also identify usability limitations in existing solutions. Studies of device fingerprinting and impersonation risk examine how contextual information can support authentication and risk assessment~\cite{alaca_device_2016,bumiller_towards_2022}. These studies motivate authentication mechanisms that account for the conditions under which access is requested.

Adaptive authentication adapts authentication methods to contextual factors such as device, location, and behavior, aiming to balance authentication requirements~\cite{hassan_engineering_2021,arias-cabarcos_survey_2020,bumiller_understanding_2023}. Risk-based adaptive authentication (RBA) research aims to assign a risk score to login attempts and adapt authentication strength accordingly \cite{Wiefling_Pump_2022}. Hulsebosch et al. \cite{hulsebosch2007context} adapt authentication methods based on location evidence fused from multiple sensors (e.g., RFID badge, Bluetooth device) associated with users, and Gebrie and Abie \cite{gebrie_risk-based_2017} use a Na\"ive Bayes classifier  in a smart-home healthcare setting to evaluate contextual risk in real time. Papaioannou et al.~\cite{papaioannou_risk-based_2021} address security and usability in authentication for mobile devices used by border-control officers, including adaptation throughout a user's interaction with the device. Dasu et al.~\cite{dasu_defending_2023} use weighted risk scoring based on factors including device and location. %

Contextual evidence also supports probabilistic and requirements-based reasoning about authentication. Hayashi et al.~\cite{hayashi_casa_2013} use probabilistic inference over contextual evidence, including location and recent activity, to select authentication challenges. Hassan et al.~\cite{10.1145/3828751} combine contextual goal models, feature models, and fuzzy causal reasoning to select authentication methods that balance security, usability, and performance. Their model already estimates attack likelihood after accounting for the mitigation provided by authentication methods and combines this likelihood with attack consequences. This framework forms the authentication phase of our approach. We extend it with an authorization phase that uses the risks remaining after authentication to determine permission restrictions and explicitly accounts for the functionality lost through those restrictions.

\paragraph{Continuous and step-up authentication}
Continuous authentication extends identity verification beyond the initial login by reassessing identity assurance during a session~\cite{mosenia_caba_2017,feng_continuous_2017}.  CABA \cite{mosenia_caba_2017} uses wearable biosignals, such as heart rate, blood pressure, and respiratory rate, using SVM and AdaBoost classifiers. It also proposes a real-time adaptive authorization scheme in which successive authentication decisions update a trust level that determines access to applications with different sensitivity thresholds. CABA therefore provides an explicit connection between continuous authentication and adaptive authorization. VAuth~\cite{feng_continuous_2017} uses a wearable accelerometer and an SVM classifier to determine whether microphone-recorded voice commands match vibrations produced by the wearer. Its classifier matches the two signals rather than identifying a particular speaker. %

Step-up authentication allows a resource server to request stronger or more recent authentication before serving a sensitive request~\cite{rfc9470}. Similarly, NIST digital identity guidelines define authentication assurance levels with progressively stronger requirements for authenticators and authentication processes~\cite{nist_sp800_63b4}. These mechanisms support decisions about the assurance required for access. Our approach focuses on authorization adaptation after a context-dependent authentication method has been selected: it relates alternative permission restrictions to attack-specific residual risk and the functionality loss that those restrictions cause.

\paragraph{Access-control models and adaptive authorization.}
Role-Based Access Control (RBAC) associates permissions with roles and assigns users to those roles, simplifying permission administration~\cite{sandhu2000nist}. Attribute-Based Access Control (ABAC) evaluates combinations of subject, object, action, and environment attributes. Standard ABAC architectures separate policy decision from enforcement and obtain relevant attributes from supporting sources~\cite{hu2014guide}. Our model takes the permissions granted by the existing authorization policy as its baseline and searches for restrictions within that baseline.
Several approaches support authorization decisions that evolve during system operation. Usage control extends traditional access control with decision continuity and mutable attributes, allowing permissions to be reassessed and revoked while usage is ongoing~\cite{10.1145/984334.984339}. Bailey et al.~\cite{bailey_self-adaptive_2011} use a feedback loop to monitor authorization usage and adapt RBAC and ABAC infrastructures in response to deviations from expected behavior. Their framework generates alternative adaptations, compares their organizational costs with the cost of doing nothing, and selects an appropriate response. It therefore already accounts for the disruption caused by restricting access. Zero-trust architectures similarly emphasize access decisions for individual resources on a per-session basis, informed by dynamic policies and contextual information~\cite{nist_sp800_207}. These approaches provide foundations for adapting authorization over time. Our work builds on these foundations by connecting permission restrictions to a formal model of residual authentication risk, operation criticality, and functionality loss.

\paragraph{Risk-adaptive access control.}

Risk-adaptive access control is closely related to our work because it incorporates security risk and operational need into authorization decisions~\cite{singhal2007secure}. Fuzzy MLS~\cite{cheng_fuzzy_2007} quantifies disclosure risk using the probability of unauthorized disclosure and the value of the information exposed, and uses this estimate to guide access decisions. NIST's account of Risk-Adaptive Access Control (RAdAC) identifies authentication-method strength, session assurance, physical location, and mission need among the factors relevant to such decisions~\cite{singhal2007secure}. Kandala et al.~\cite{kandala2011radac} develop an attribute-based RAdAC framework that represents operational needs, risk, and situational factors. Their framework includes mission and operation attributes, allowing different operations on the same information to carry different risks. In healthcare, Wang and Jin~\cite{wang2011quantified} quantify the risk associated with physicians' data-access activities to detect and control over-access to patient information while accommodating exceptional information needs. Armando et al.~\cite{debruyne_balancing_2015} combine trust and risk in an access-control framework that grants access when trust meets or exceeds risk. Their model explicitly accounts for authentication uncertainty through an identity-theft probability associated with authentication strength. It also considers residual risk after security controls, though not the risk specific to a chosen authentication method, and supports further risk mitigation, such as data obfuscation, or trust enhancement through enforceable obligations. Their approach thus already connects authentication uncertainty with authorization. However, authentication strength enters only as a single scalar that scales the requester's trust, and the authors identify the systematic study of how authentication mechanisms affect risk-aware access control as future work.

These studies establish that authorization can account for authentication assurance, operational importance, and the consequences of restricting access. Our contribution lies in combining these concerns in a decision model for authorization adaptation following adaptive authentication. The model estimates residual risk for each modeled attack under the selected authentication method and current context, represents operation criticality and asset-permission dependencies, and assigns penalties to permission restrictions and operation disablement. It then identifies alternative authorization configurations with explicit residual-risk and functionality trade-offs, retaining the access required by Must-Have operations and supporting administrative selection among the resulting configurations.

\paragraph{Emergency access.}
Break-glass mechanisms allow users to override access-control decisions in exceptional situations through controlled and accountable procedures~\cite{brucker_extending_2009}. Such mechanisms can expand access when urgent legitimate needs are not met by the current policy. Our approach restricts access to reduce residual risk while retaining the permissions required by operations classified as Must-Have, including designated emergency operations whose asset and permission dependencies are captured in the model. A separate break-glass mechanism could support exceptional access needs not captured by this classification. This is a possible complementary mechanism rather than a feature implemented by our approach, as discussed in Section~\ref{threat_to_validaty}.

\paragraph{Residual risk.}

In security risk management, residual risk denotes the risk that remains after security controls are applied~\cite{nist_sp800_30r1}. S\'anchez-Garc\'ia et al.~\cite{app13010395} include residual-risk assessment in their cybersecurity risk-assessment model. Anisetti et al.~\cite{9590416} integrate risk management with assurance monitoring and use flow networks to compute risk mitigation and residual risk in distributed systems. These contributions provide methods for assessing the risk remaining after controls and monitoring their effectiveness. We apply residual-risk reasoning to a sequence of security decisions: authentication reduces the likelihood of modeled attacks, while subsequent authorization restrictions reduce their potential impact through changes to the assets and actions accessible to the session.

Prior work provides mechanisms for adapting authentication and authorization, incorporating operational needs, and assessing residual risk. Our approach brings these elements into a formal decision problem that links the authentication method selected in the current context to the authorization configurations available afterward. Its focus is on quantifying how alternative restrictions affect both residual risk and functionality, while preserving the modeled access requirements of the most critical operations.

%% file: Conclusion.tex
\section{Conclusion and Future Work}

 Selecting an adaptive authentication method does not eliminate attack risks, whose consequences also depend on the permissions available to a compromised session. This paper presents adaptive authorization as a second line of defense by quantifying residual risk and recommending feasible authorization configurations that reduce it while preserving essential functionality. The approach models operation criticality, asset-operation dependencies, fine-grained access rights, and functionality-loss penalties, using SMT-LIB and Z3 to reason about risk–functionality trade-offs.

Evaluation across healthcare and industrial control system scenarios shows that the approach can reduce overall and attack-specific residual risks while preserving Must-Have operations. Further risk reduction may require tighter access rights and disabling lower-priority operations. The resulting configurations allow administrators to select restrictions according to operational priorities and risk appetite. Baseline comparisons and robustness analysis support the approach's effectiveness within the evaluated scenarios, while runtime and memory measurements quantify its computational overhead.

The findings depend on the accuracy of risk estimates, modeled dependencies, and manually assigned operation criticality. Future work will evaluate a wider range of authentication attacks and contextual conditions, investigate reinforcement learning for adjusting contextual weights and modeled authentication effects under expert validation and Must-Have constraints, and explore historical access logs and operational evidence to inform operation classification, particularly when usage frequency alone does not indicate criticality.